\documentclass[useAMS,usenatbib]{mn2e}

\usepackage{graphicx}
\usepackage{txfonts}
\usepackage{natbib}
\usepackage{longtable,lscape}
\usepackage{color}
\usepackage[T1]{fontenc}
\usepackage{aecompl}

\newcommand\aj{AJ}
\newcommand\apj{ApJ}
\newcommand\apjs{ApJS}

\newcommand\aap{A\&A}
\newcommand\mnras{MNRAS}
\newcommand\apjl{ApJ}
\newcommand\pasp{PASP}

\newcommand\aaps{A\&AS}
\newcommand\araa{ARA\&A}

\def\teff{\mbox{T$_{\rm eff}$}}
\def\logg{\mbox{log~{\it g}}}
\def\vmicro{\mbox{$\xi_{\rm t}$}}
\def\kmsec{\mbox{km~s$^{\rm -1}$}}

\title[]{Chemical abundances in the multiple sub-giant branch of 47~Tucanae: insights on its faint sub-giant branch component}
\author[A.\, F.\, Marino et al.]
{A.\, F.\, Marino$^{1}$,
A.\,P.\, Milone$^{1}$,
L. Casagrande$^{1}$,
R. Collet$^{1,2}$,
A. Dotter$^{1}$,
C.\,I.\, Johnson$^{3}$,
\newauthor
K. Lind$^{4}$,
L.\,R.\, Bedin$^{5}$,
H. Jerjen$^{1}$,
A. Aparicio$^{6}$,
L. Sbordone$^{7,8}$
\\
$^{1}$Research School of Astronomy \& Astrophysics, Australian National University, Canberra, ACT 2611, Australia \\
$^{2}$Stellar Astrophysics Centre, Department of Physics and Astronomy, Aarhus University, Ny Munkegade 120, DK-8000 Aarhus C, Denmark \\
$^{3}$Harvard–Smithsonian Center for Astrophysics, 60 Garden Street, MS-15, Cambridge, MA 02138, USA \\
$^{4}$Max-Planck-Institut fuer Astronomie, Koenigstuhl 17, D-69117 Heidelberg, Germany\\
$^{5}$INAF-Osservatorio Astronomico di Padova, Vicolo dell’Osservatorio 5, I-35122 Padova, Italy\\
$^{6}$Instituto de Astrof\'isica de Canarias, La Laguna, Tenerife, Spain\\
$^{7}$Department of Electrical Engineering, Center for Astro-Engineering, Pontificia Universidad Cat\'{o}lica de Chile, Av. Vicu\~{n}a Mackenna 4860, 782-0436 Macul, Santiago, Chile\\
$^{8}$The Milky Way Millennium Nucleus, Av. Vicu\~{n}a Mackenna 4860, 782-0436 Macul, Santiago, Chile\\
}

\begin{document}

\date{Draft Version Mar, 2016}

\pagerange{\pageref{firstpage}--\pageref{lastpage}} \pubyear{2016}

\maketitle

\label{firstpage}

\begin{abstract}  
The globular cluster 47~Tuc exhibits a complex sub-giant branch (SGB) with a faint-SGB comprising only about the 10\% of the cluster mass and a bright-SGB hosting at least two distinct populations.
We present a spectroscopic analysis of 62 SGB stars including 21 faint-SGB stars. We thus provide the first chemical analysis of the intriguing faint-SGB population and compare its abundances with those of the dominant populations.
We have inferred abundances of Fe, representative light elements C, N, Na, and Al, $\alpha$ elements Mg and Si for individual stars. Oxygen has been obtained by co-adding spectra of stars on different sequences. In addition, we have analysed 12 stars along the two main RGBs of 47~Tuc.

Our principal results are: (i) star-to-star variations in C/N/Na among RGB and bright-SGB stars; (ii) substantial N and Na enhancements for the minor population corresponding to the faint-SGB; (iii) no high enrichment in C$+$N$+$O for faint-SGB stars. Specifically, the C$+$N$+$O of the faint-SGB is a factor of 1.1 higher than the bright-SGB, which, considering random ($\pm$1.3) plus systematic errors ($\pm$0.3), means that their C$+$N$+$O is consistent within observational uncertainties. However, a small C$+$N$+$O enrichment for the faint-SGB, similar to what predicted on theoretical ground, cannot be excluded. 
The N and Na enrichment of the faint-SGB qualitatively agrees with this population possibly being He-enhanced, as suggested by theory.
The iron abundance of the bright and faint-SGB is the same to a level of $\sim$0.10~dex, and no other significant difference for the analysed elements has been detected.
\end{abstract}

\begin{keywords}
globular clusters: general -- individual: NGC\,104 (47~Tucanae) -- techniques: spectroscopy
\end{keywords}

\section{Introduction}\label{sec:intro}

The presence of multiple stellar populations in globular clusters (GCs) has been
widely demonstrated both photometrically and spectroscopically (see Piotto et al.\,2015 and references therein). {\it Hubble Space Telescope} ({\it HST\,}) photometry has revealed that the phenomenon of multiple sequences along the color-magnitude diagrams (CMDs) is not exclusive to $\omega$~Centauri. Many chemical abundance studies from different groups have shown the presence of intrinsic variations in $p$-capture elements (e.g., Kraft 1994; Ivans et al.\,1999; Gratton et al.\,2001; Marino et al.\,2008).
Evidence for C-N, O-Na (and in some cases Mg-Al) anticorrelations among GC stars suggest that the pollution from high-mass asymptotic giant branch stars (AGB) and/or fast-rotating massive stars has occurred in a self-pollution scenario (e.g., Ventura et al.\,2001; Decressin et al.\,2007), or alternatively from early accretion disks in pre-main-sequence binary systems (Bastian et al.\,2013, see Renzini et al.\,2015 for a critical discussion on the different scenarios). 

A few GCs, so-called {\it anomalous} GCs, are known to have intrinsic metallicity variations (see Tab.~10 in Marino et al.\,2015, and reference therein). On the photometric side, these GCs show multiple sub-giant branches (SGBs). While the presence of multiple main sequences (MSs) and red-giant branches (RGBs) in UV bands, as well as in some cases extended horizontal branches (HBs), are in general good indicators for variations in light elements (including He for the HB; see Milone et al.\,2014), multiple SGBs are observed in many of the clusters with Fe variations independently on the photometric band (e.g., Milone et al.\,2008; Piotto et al.\,2012). 

Many attempts have been made in the past years to connect the CMD multiple sequences with the GC chemical variations in order to understand how successive stellar generations could have formed (e.g., Marino et al.\,2008; Yong et al.\,2008; Milone et al.\,2012a). 
While the RGB-chemistry connection has been assessed through chemical abundances studies of bright RGB stars, the chemistry of multiple MSs and SGBs remains more unexplored. Evolutionary models suggest that the appearance of multiple MSs is consistent with stellar populations of different He (e.g., Bedin et al.\,2004; Norris 2004; D'Antona et al. 2005). 
Given that the nucleosynthesis processes responsible for producing the N/Na/Al enhanced and C/O depleted stars in GCs are also expected to produce He, connecting multiple photometric sequences with varying levels of He enrichment is a reasonable assumption.

Instead of what was predicted for the MS, He differences cannot account for the size of the observed SGB splits in {\it anomalous} GCs (Milone et al.\,2008; Piotto et al.\,2012), that could be explained either by age variations among stellar populations and/or by differences in the overall C$+$N$+$O (Cassisi et al.\,2008). In support of the latter scenario, variations in the total CNO were found among RGB stars of NGC\,1851 (Yong et al.\,2009, 2015), M\,22 (Marino et al.\,2011), and $\omega$~Centauri (Marino et al.\,2012a). 
Spectroscopic studies have been conducted on SGB stars in M\,22 and NGC\,1851, showing that fainter SGB stars are more enriched in Fe and $s$-elements than brighter ones, consistent with being also C$+$N$+$O (Marino et al.\,2012b; Gratton et al.\,2012). 

High-accuracy photometry from the $HST$ has revealed a double MS in NGC\,104 (47~Tuc, Milone et al.\,2012a). Multiple populations on the SGB were identified by Anderson et al. (2009) who found at least two distinct branches: a brighter SGB with an intrinsic broadening in luminosity, and a fainter one populated by a small fraction of stars. 
Theoretical models by Di Criscienzo et al.\,(2010) can reproduce the lower luminosity of the less populous SGB only by assuming either a C$+$N$+$O overall enhancement, or an age difference of about 1~Gyr.

Early spectroscopic studies have shown that 47~Tuc is bimodal in the distribution of the CN bands strengths (e.g., Norris \& Freeman\,1979; Cannon et al.\,1998), but no evidence for a variation in the total C$+$N$+$O were revealed. However, stars that are predicted to be CNO enhanced constitute only 10\% of the SGB stars, and none of them is included in previous studies on 47~Tuc subgiants (e.g., Carretta et al.\,2005).
Furthermore, 47~Tuc is not among the newly defined class of {\it anomalous} GCs, e.g., clusters with clear internal variations in $s$-process elements and iron.
Only one star has been found to be enhanced in the $s$-process composition, but it is interpreted as due to binarity (Cordero et al.\,2015).

In this paper we explore the chemical composition of different photometric sequences along the SGB and the RGB of 47~Tuc, from the MS up to the low RGB, with more focus on the SGB sub-populations. 
The paper is organised as follows: Sect.~\ref{sec:data} presents the photometric and spectroscopic data that we used; Sect.~\ref{sec:atm} explains how atmospheric parameters have been derived, and Sect.~\ref{sec:abundances} the chemical abundances we were able to infer; our results are presented and discussed in Sect.~\ref{sec:abb}  and Sect.~\ref{sec:sgb}; Sect.~\ref{sec:conclusions} is a summary of the main results.

\section{Data}\label{sec:data}

\begin{figure*}
   \centering
   \includegraphics[width=11.0cm]{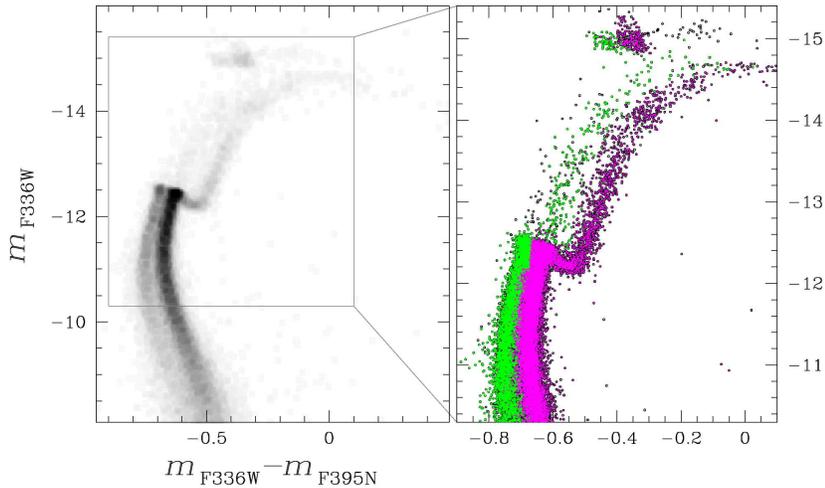}
      \caption{$m_{\rm F336W}$ vs.\,$m_{\rm F336W}-m_{\rm F395N}$ Hess diagram of 47\,Tuc from UVIS/WFC3 photometry (left panel). Right panels is a zoom of the CMD in the region where the two main populations of 47\,Tuc, namely A and B, have been colored green and magenta, respectively. }
        \label{fig:CMDhst}
   \end{figure*}

Recent studies, based on both ground-based and {\it HST} photometry have shown that 47\,Tuc hosts two main populations, that can be identified in appropriate CMDs and colour-colour diagrams, traced continuously along the MS, the SGB, the RGB and the HB (Milone et al.\,2012a). 
The two most prominent populations of 47\,Tuc are clearly visible in the $m_{\rm F336W}$ vs.\,$m_{\rm F336W}-m_{\rm F395N}$ Hess diagram plotted in the left panel of Fig.~\ref{fig:CMDhst}. In the CMD shown in the right panel of Fig.~\ref{fig:CMDhst} we have used the green and magenta colours to represent MS, SGB, RGB, and HB stars of the population-A and population-B stars identified by Milone et al.\,(2012a), respectively.
Spectroscopic and photometric studies have shown that population-A is made of stars with primordial helium, high oxygen and low sodium and includes $\sim$20\% of all stars in the cluster core, while the four times larger population-B is enhanced in He and Na and depleted in O (see Milone et al.\,2012a). 
 
A third population has been detected along the SGB-B only. It is distributed at fainter magnitudes and includes only the $\sim$10\% of the total number of cluster stars (Anderson et al.\,2009; Piotto et al.\,2012). Its chemical composition is still unknown. 
In the following subsections we will use both photometry and spectroscopy to characterize the multiple stellar populations of 47\,Tuc along the SGB and the RGB, focusing on the minor SGB steller population detected by Anderson et al.\,(2009) and Piotto et al.\,(2012).

%
%

\subsection{The photometric dataset\label{sec:phot_data}}

 The photometric dataset used in our study of 47~Tuc consists of {\it HST} photometry for the innermost $\sim$3$\times$3 arcmins and ground-based photometry for the outer cluster region.
 Specifically, we have used the catalogs published by Milone et al.\,(2012a) that include photometry in the F336W (30s+1160s images from GO\,11729, PI.\,Holtzmann), F395N bands (90s+2$\times$1120s images from GO\,11729, PI.\,Holtzmann) from the Ultraviolet and Visual Channel of the Advanced Camera for Surveys 3 (UVIS/WFC3) and in the F435W (9$\times$105s images from GO\,9281, PI.\,Grindlay), F606W (3s$+$4$\times$50s images from GO\,10775, PI.\,Sarajedini), and F814W (3s$+$4$\times$50s images from GO\,10775, PI.\,Sarajedini) band from the Wide Field Channel of the Advanced Camera for Surveys (WFC/ACS)  on board of {\it HST}.
We refer to the paper by Milone and collaborators and references therein for details on the data and the data reductions.

\begin{figure*}
   \centering
   \includegraphics[width=9.0cm]{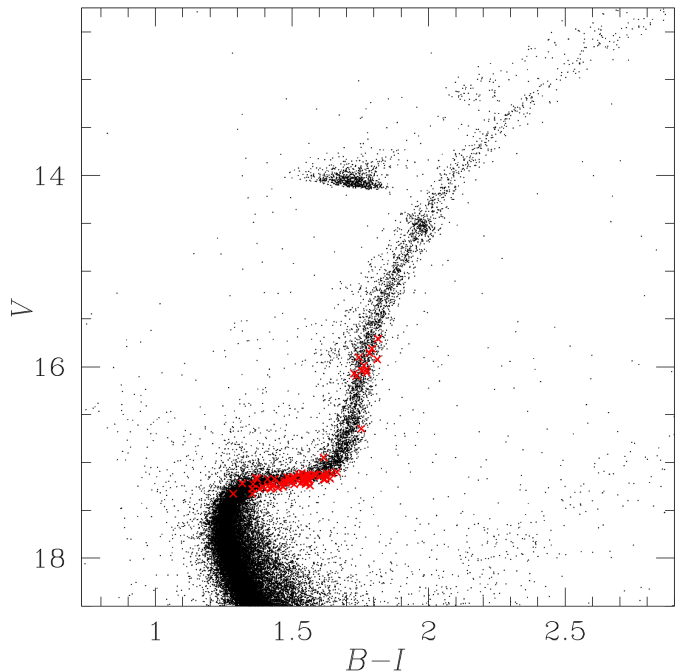}
      \caption{
$V$ vs.\,$B-I$ CMD from ground-based photometry (from Stetson\,2000, and Bergbush \& Stetson\,2009). The spectroscopic targets analyzed in this paper are indicated with red crosses.}
        \label{fig:cmdtargets}
   \end{figure*}

 Moreover, we have used ground-based photometry in the {\it U}, {\it B}, {\it V}, and {\it I} filters from the archive maintained by Peter Stetson. The photometric catalog has been obtained  by Stetson (2000) from 856 CCD images, including 480 images taken with the Wide-Field Imager
of the ESO/MPI 2.2 m telescope, and 200 with the 1.5 m telescope at
Cerro Tololo Inter-American Observatory. The remaining 176 images
come from various other telescopes.  All the images were reduced using the method described by Stetson (2005) and are calibrated on the Landolt (1992) photometric system. We refer to the paper by Bergbusch \& Stetson (2009) for further details on this dataset.
The CMDs from {\it HST} and ground-based photometry are plotted in Fig.~\ref{fig:CMDhst} and Fig.~\ref{fig:cmdtargets} respectively, where the spectroscopic targets are marked with red crosses in Fig.~\ref{fig:cmdtargets}.

Ground-based photometry has been used to select the targets, derive the atmospheric parameters and to identify multiple populations along the RGB and the SGB. {\it HST} photometry has been exploited to illustrate the multiple populations of this cluster and to constrain the age and the overall CNO abundance of the faint and the bright SGB by comparing the CMD with appropriate isochrones. 

%
%

%
   \begin{figure*}
   \centering
   \includegraphics[width=8.8cm]{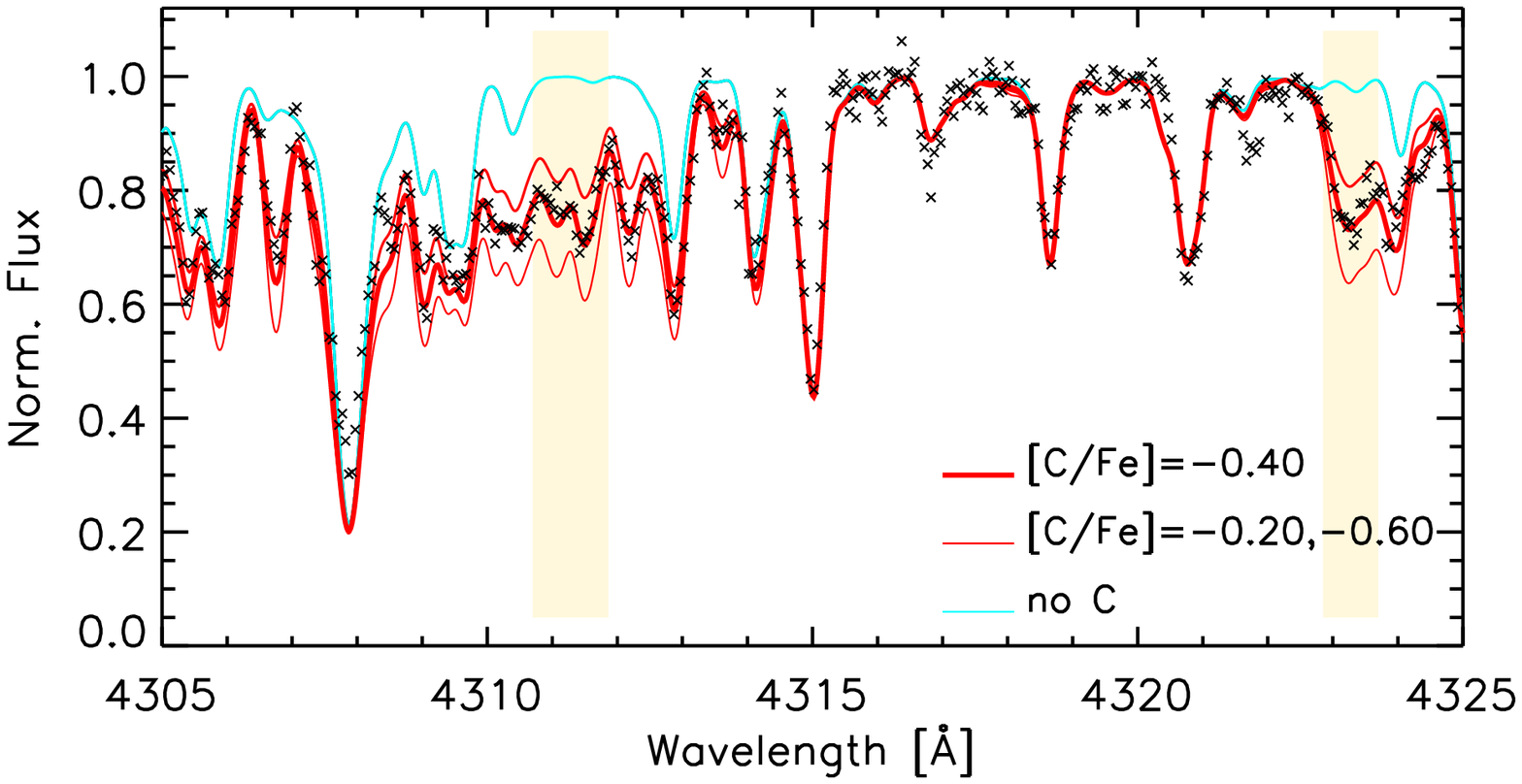}
   \includegraphics[width=8.8cm]{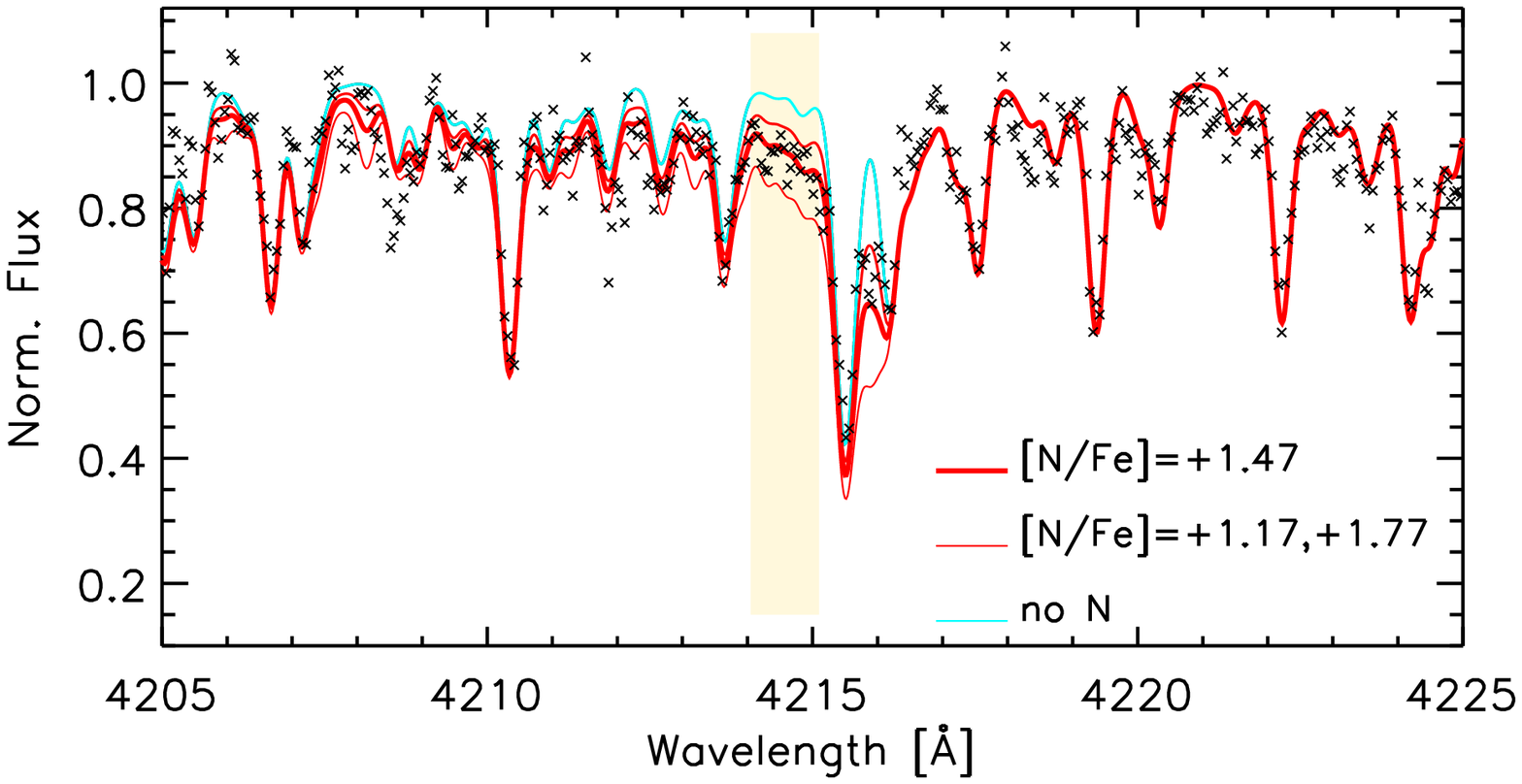}
      \caption{Observed spectra for the SGB star N104e-5623 around the CH (left panel) and CN (right panel) molecular bands. For each band, we represent the best-fit synthesis (solid-bold red line), the synthesis with [C/Fe] and [N/Fe] varied by 0.20~dex and 0.30~dex for CH and CN (solid-light red lines), respectively, and the synthesis with no C or N (aqua line). The spectral regions used to derive the chemical abundances have been shown with shaded regions. }
        \label{fig:spectra}
   \end{figure*}
%

\subsection{The spectroscopic dataset}\label{sec:spec_data}

Our spectroscopic data consist of FLAMES/GIRAFFE spectra (Pasquini et al.\ 2002)\nocite{pas02} observed under the program 089.D-0579(A) (PI.\,Marino).
The HR02, HR04 and HR19A GIRAFFE setups were employed, which cover spectral ranges 
$\sim$3854-4049~\AA, $\sim$4188-4392~\AA, and
$\sim$7745-8335~\AA, and have resolving powers $R \equiv \lambda/\Delta\lambda \sim$19,600-20,350-13,900, respectively.

Our sample of stars, which is represented in Fig.~\ref{fig:cmdtargets}, has been accurately selected to sample the different
SGBs and RGBs of 47~Tuc and is made up of 12 RGB stars, with 15.5$\lesssim V\lesssim$17.0 and 62 SGB stars with 17.0$\lesssim V\lesssim$17.5. All the targets were observed with the same FLAMES plate in 3-13-17 different exposures of 46 minutes for HR02, HR04, and HR19A, respectively. 
The typical S/N of the fully reduced and combined GIRAFFE spectra
ranges from $\sim$30 to $\sim$60 at 4300~\AA\ (HR04),
$\sim$50 to $\sim$130 at 8180~\AA\ (HR19A), depending on the
brightness of the star.  
Given the lower priority given to the HR02 setup, not all our requested exposures were executed and we could not reach S/N higher than 10, and we decided to not use this setup. 

Data reduction involving bias-subtraction, flat-field correction, 
wavelength-calibration, sky-subtraction, continuum normalisation,
has been done by using the dedicated pipelines\footnote{See {\sf
    http://girbld-rs.sourceforge.net}} and IRAF standard procedures. 
Telluric absorptions strongly affect the H19A setup, in particular in the range 8100-8335\AA. We have removed tellurics by using the software MOLECFIT\footnote{{\sf
    http://www.eso.org/sci/software/pipelines/skytools/molecfit}} (Smette et al.\,2015; Kausch et al.\,2014). But, even with such procedure, we caution that residual telluric
feature contamination might be still of concern for the analysis of the
spectral lines in the contaminated spectral region. 

Radial velocities (RVs) were derived separately from each setup using the IRAF@FXCOR task, which cross-correlates the object spectrum with a template. For the template we used a synthetic spectrum obtained through MOOG (Sneden C.\,1973, version February, 2014\footnote{\sf http://www.as.utexas.edu/~chris/moog.html}). 
This spectrum was computed with a model stellar atmosphere interpolated from the Castelli \& Kurucz (2004)\nocite{kur92} grid\footnote{{\sf http://kurucz.harvard.edu/grids.html}}, adopting parameters (\teff, \logg, \vmicro, [Fe/H]) = (5600~K, 3.8, 1.0~\kmsec, $-$0.72). Each spectrum was corrected to the restframe system, and observed RVs were then corrected to the heliocentric system.
We do not find strong systematic differences among the average RVs obtained from the three different setups, being the mean offsets RV$_{\rm HR19} -$RV$_{\rm HR4}$=$+0.5\pm0.1$~\kmsec\ and RV$_{\rm HR19} -$RV$_{\rm HR2}$=$-0.2\pm0.9$~\kmsec.
The final RVs listed in Tab.~\ref{tab:data} were computed from the weighted mean of the RVs from the HR4 and HR19 setups, neglecting the HR2 data which give higher internal uncertainties.
Deriving the RV also served as an independent check of cluster membership. We find a mean RV of $-16.73\pm0.77$~\kmsec, with rms=6.66~\kmsec, consistent with the literature results (Alves-Brito et al.\,2005; Koch \& McWilliam 2008; Dobrovolskas et al.\,2014).
Coordinates, basic $BV$ photometry and RVs for the all the analysed stars are listed in Tab.~\ref{tab:data}. 

\section{Model atmospheres}\label{sec:atm}

The spectral range covered by the HR04+HR19A setups provides a relatively low number of Fe lines to ensure an optimal determination of atmospheric parameters. Furthermore, at the relatively high-metallicity of 47~Tuc ([Fe/H]=$-$0.72, Harris\,1996, 2010), the HR04 spectral range is crowded and most of the lines are blended.

We interpolated isochrones from Dotter et al.\,(2008), and used a E$(B-V)$=0.025, $(m-M)_{V}$=13.42, in agreement with values from Harris\,(1996, updated as in 2010), and age=12.0~Gyr. 
In the subsequent analysis we adopted photometry corrected for differential reddening (see Milone et al.\,2012b for details on the procedure) plus a mean reddening of E$(B- V)$=0.025 in all instances.
We find that the average reddening variation of stars in the spectroscopic sample is E$(B-V)$=0.003, with the minimum and maximum values corresponding to E$(B-V)$=$-$0.007 and E$(B-V)$=$+$0.009, respectively.

Given the difficulties in the determination of the atmospheric parameters from spectral lines, we take advantage of our $BV$ photometry (see Sect.~\ref{sec:phot_data}). For the SGB stars, we derive effective temperatures (\teff) from the $(B-V)$-\teff\ relation provided in Casagrande et al.\,(2010), which is suitable for dwarfs and sub-giants with surface gravities (\logg) higher than 3.5.
Then, we compared these \teff\ values with the ones obtained from the isochrone that best fits our CMD. 
We obtained a mean difference among the two sets of temperatures of \teff$_{\rm isochrone} -$\teff$_{\rm (B-V)}$=$-33 \pm 3$~K for sub-giants, with rms $< 30$~K. This spread suggests that the internal uncertainties associated with the photometric \teff\ values is likely small, so we assume an error of 50~K. 

For red giants $JHK$ magnitudes from 2MASS are more accurate than for subgiants, so that for the former stars we could apply the IRFM directly, by using $BVIJHK$ photometry (Casagrande et al.\,2010, 2014)
In this case, the mean difference with the \teff\ values given by the best-fit isochrone is \teff$_{\rm isochrone} -$\teff$_{\rm (IRFM)}$=$-51 \pm 26$~K, with rms of 82~K.
Our \teff\ values were then corrected to the best-fit isochrone scale, which means we added to the IRFM and $(B-V)$ colour-based temperatures the offsets calculated for the giants and sub-giants.

Surface gravities were obtained from the apparent $V$ magnitudes, corrected for differential reddening, the \teff\ from above, bolometric corrections from Alonso et al. (1999), and an apparent distance modulus of $(m-M)_{V}$=13.42. We assume masses taken from the best-fit isochrone, which range from 0.81 to 0.88~${\rm M_{\odot}}$. 
We notice that our adopted masses agree with the empirical estimate of the mass of turn-off stars in 47~Tuc measured from an eclipsing binary, which is 0.8762~$M_{\odot}$ for the primary and 0.8588~$M_{\odot}$ for the secondary (Thompson et al.\,2010).

For microturbulent velocities (\vmicro), we adopted the appropriate relation used in the Gaia-ESO survey (GES, Gilmore et al.\,2012) 
\footnote{For clarity, we used: 
$\vmicro=0.94+2.2e-5\times(\teff-5500)-0.5e-7\times((\teff-5500)^{2})-0.1\times(\logg-4.00)+0.04\times((\logg-4.00)^{2})-0.37\times[Fe/H]-0.07\times[Fe/H]^{2}$ for giants (\teff$<$5250 and \logg$<$3.5); 
and 
$\vmicro=1.15+2e-4\times(\teff-5500)+3.95e-7\times(\teff-5500)^{2}-0.13\times(\logg-4.00)+0.13\times(\logg-4.00)^{2}$ for MS and SGB stars (\teff$\geq$5250 and \logg$\geq$3.5).
}, 
that depends on \teff, \logg\ and metallicity.
Temperatures and gravities were already set from above, while for the metallicity we assumed [A/H]=$-$0.72 (Harris 1996, as updated in 2010).
For both \logg\ and \vmicro\ we assumed an internal error of 0.20.

\section{Chemical abundances analysis}\label{sec:abundances}

Chemical abundances were derived from a local thermodynamic equilibrium (LTE) analysis by using the line analysis code MOOG (Sneden\,1973, version Feb. 2014). We also used the $\alpha$-enhanced model atmospheres of Castelli \& Kurucz\,(2004) and the parameters described in Sect.~\ref{sec:atm}.
Reference solar abundances are from Asplund et al.\,(2009). 

{\it Elements involved in proton-capture reactions:}
Carbon was inferred from spectral synthesis of the CH G-band $(A^{2}\Delta - X^{2}\Pi)$ heads near 4312 and 4323~\AA. Nitrogen was derived from synthesis of the CN blue system $(B^{2}\Sigma - X^{2}\Sigma)$ bandhead at  $\sim$4215~\AA. The synthesis linelists for the CH G band are from Masseron et al.\,(2014), while for the blue CN band we used the Kurucz linelist\footnote{{\sf http://kurucz.harvard.edu/linelists.html}}. As an example of the molecular band synthesis we show synthetic/observed spectral matches of the CH and CN for the SGB star N104e-5623 in Fig.~\ref{fig:spectra}. Superimposed on the observed spectra are best-fit synthetic models, synthetic models where C and N have been varied by 0.2 and 0.3, respectively, and models with no C and N. 

We determined Na abundances from the doublet at $\sim$8190\AA, Mg from the line  at $\sim$8213~\AA, and the features at $\sim$7931 and 8099~\AA. Aluminum was inferred from the doublet at $\sim$7836~\AA. 
The Na abundances were corrected for NLTE effects adopting the corrections from Lind et al.\,(2011).
For oxygen we analysed the triplet at $\sim$7770~\AA.

{\it Heavy iron-peak elements:}
Ni abundances have been inferred from the line at $\sim$7798~\AA\, detectable only in our RGB sample.

{\it $\alpha$ elements:}
For the $\alpha$~elements we determined only abundances of Si from a single line at $\sim$7944~\AA. We note that silicon can also be affected by $p$-capture, as well as Mg, which has been commented with the other $p$-capture elements. 

Internal uncertainties in chemical abundances due to the adopted model atmospheres were estimated by varying the stellar parameters, one at a time, by the amounts discussed in Sect.~\ref{sec:atm}, namely 
$\pm$50~K/$\pm$0.20~dex/$\pm$0.05~dex/$\pm$0.20~\kmsec.
The impact on the chemical abundances due to changes in the various stellar parameters are listed in Tab.~\ref{tab:errGIR}.
The limited S/N of our spectra introduces significative internal uncertainties to our chemical abundances. 
To estimate these uncertainties in the spectral synthesis we computed a set of 100 synthetic spectra for 
two stars representative of the sample (the SGB star N104e-51475 and the RGB star N104e-31450). These set of synthetic spectra were calculated by using the best-fit inferred abundances, and were then degraded to the S/N of the observed spectra. We then analysed the chemical abundances of all these synthetic spectra in the same manner as the observed spectra. The scatter in the abundance measurements that we obtain from the 100 synthetic spectra corresponding to a given star, is a fair estimate of the uncertainty introduced by the fitting procedure, due to the S/N, the pixel size and the continuum estimate. 
These error estimates are listed in Tab.~\ref{tab:errGIR} as $\sigma_{\rm fit}$, and, of course, are higher for SGB stars than RGB stars and for elements whose abundance has been inferred by one or two lines only. 
The last column of Tab.~\ref{tab:errGIR} lists the total internal errors in our chemical abundances, due to the stellar parameters and the quality of data ($\sigma_{\rm tot}$).
Other contributions to the errors are due to effects of C and N, so the $\sigma_{\rm tot}$ values are an underestimation of the real error (see Sect.~\ref{CNeffects}).

Systematic uncertainties are likely  higher, and may be introduced by various sources, such as: 
{\it (i)} systematics in atmospheric parameters;
{\it (ii)} adopted atomic data;
{\it (iii)} possible 3D effects in the case of molecular bands;
{\it (iv)} NLTE effects.
Most of these effects are likely to affect systematically the abundances of all the stars with similar parameters. So, they are not expected to strongly impact on our results based on star-to-star internal variations. Some of these effects may affect stars at various evolutionary stages, e.g.,  SGB and RGB stars, in a different manner. 
We will discuss these effects below, when we judge them to possibly be relevant.

\subsection{The effect of C and N}\label{CNeffects} 

One of the most serious effects that might alter our results is the presence of molecular bands in the HR19A GIRAFFE setup, which is affected by CN features. 
The analysis of star-to-star internal chemical variations would not be affected if the analysed stars have the same CN abundances, as, in that case, all the abundances would be only systematically shifted. However, in the case of a GC, the different abundances of C and N can potentially affect the results in a different way for the first and second population stars, that have different C and N abundances.

To evaluate if and how CN molecular bands, present around $\sim$8000~\AA, affect our results for abundances other than CN, we ran the spectral synthesis of each analysed spectral feature by using different C and N compositions. Specifically, we used the average composition for stars in different locations of our observed C-N anticorrelation, which will be discussed in Sect.~\ref{sec:abb}. For the SGB stars we chose two different sets of compositions for C and N ([C/Fe], [N/Fe])=($-$0.2,$+$0.6),($-$0.4,$+$1.4), as well as for the RGB ([C/Fe], [N/Fe])=($-$0.10,$+$0.27),($-$0.22,$+$0.90), matching the extreme values observed on the [C/Fe]-[N/Fe] plane. 

\begin{figure}
   \centering
   \includegraphics[width=8.7cm]{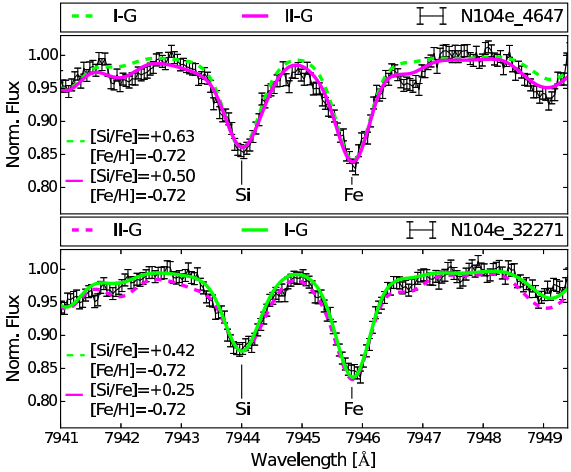}
      \caption{Observed spectra of the RGB stars N104e-4647 (upper panel) and N104e-32271 (lower panel) around the Si line at $\sim$7944~\AA. For both stars two best-fit synthesis models have been super-imposed to the observed spectra: one computed with the mean CN abundance of first-generation (I-G, green) stars (as obtained from our analysis), and the other with the mean CN abundance for the second-generation (II-G, magenta) stars. In both panels, the synthesis that we used is represented with a solid line, the other with a dashed line. The abundance of Si and Fe used in the synthesis is quoted in each panel.}
        \label{fig:SiCN}
   \end{figure}

We found that, while the Na and Fe results are almost unaffected by the different C and N composition, the effect of molecular bands is not negligible for other elements, the most affected feature being the Si line at $\sim$7944~\AA. In Fig.~\ref{fig:SiCN} we represent two observed spectra, one for a C-poorer/N-richer RGB star (N104e-4647), and the other for a C-richer/N-poorer RGB star (N104e-32271), around the Si line. Two spectral synthesis have been super-imposed to the observed spectra, one with the average C and N abundances of N-poorer stars (first generation, I-G), and the other with the average abundances for N-richer stars (second-generation, II-G). It is clear that the spectral features around the Si line are best reproduced by a I-G C and N composition in the case of N104e-32271, and by a II-G composition for N10e-4647.
The best-fit Si abundance is different depending on the C and N assumed in the synthesis: for both stars we get a different [Si/Fe], depending on what set of C and N abundances (I-G or II-G) we assume. 
The difference between the Si abundances is $\sim$0.15~dex, depending on which C and N composition we assume.

Another unaccounted effect for the RGB stars is that, since the most common C diatomic molecule is CO, the amount of C free to form CH and CN is dependent on O abundance, which we are not able to measure and has been fixed to a value of [O/Fe]=$+$0.30 for all the stars. In reality, stellar populations in GCs have different O abundances. To test how our assumption of equal O affect our results we have performed some spectral synthesis of CH and CN bands for typical SGB and RGB stars by assuming a variation in [O/Fe] of 0.5~dex, between [O/Fe]=$-$0.10 and [O/Fe]=$+$0.40, which is what is typically observed in GCs. Our synthesis suggest that this variation in [O/Fe] does not significantly affect the C and N abundances in our SGB sample as the differences between the synthetic spectra determined with [O/Fe]=$-$0.10 and [O/Fe]=$+$0.40 are undetectable. On the other hand, differences in the C abundances are larger in our sample of giants: specifically, the O-poor model requires a [C/Fe] lower by $\sim$0.1~dex from the G band; the N abundances are less affected ($\sim$0.05~dex) as the combined variations in C and O cancel out in the synthesis of the CN band.

In the spectral range represented in Fig.~\ref{fig:SiCN} falls a Fe line, which we used. Noticeably, both the synthetic spectra have the same Fe abundance, that means this line is not significantly affected by CN variations (our tests suggest that, in general, all our used lines are not significantly affected).

For SGB stars, CN variations have a smaller impact, being the mean difference in [Si/Fe] abundance between the synthesis computed by assuming a I-G or II-G composition of 0.05~dex. Our abundance analysis accounts for CN variation only for RGB stars. This means that our RGB sample, once we got guess information on C and N abundances, we run the synthesis of the Si line a second time by using the mean C and N abundances of the two RGB populations. 

The impact of CN molecules on other chemical species is smaller than that on Si. Magnesium and aluminum are mildly affected at a level of $\sim$0.05~dex. In conclusion, we can assume a tenth of dex as a realistic total internal error on Mg and Al. For Si we consider a total uncertainty of 0.15~dex, while for the other elements the $\sigma_{\rm tot}$ values listed in Tab.~\ref{tab:errGIR} are fair estimates of our internal errors.

%
%

\section{The chemical composition along the CMD}\label{sec:abb}

The mean metallicity obtained from our GIRAFFE sample of 47~Tucanae probable members, composed by 74 stars, is [Fe/H]=$-$0.73$\pm$0.01~dex, with a dispersion $\sigma$=0.09~dex. 
We do not observe any significant difference in the Fe content between RGB and SGB stars: the mean abundances for these two subsamples are [Fe/H]$_{\rm RGB}=-$0.77$\pm$0.02 (rms=0.08) and [Fe/H]$_{\rm SGB}=-$0.72$\pm$0.01 (rms=0.09), which are the same within 2~$\sigma$.

\begin{figure}
   \centering
   \includegraphics[width=8.0cm]{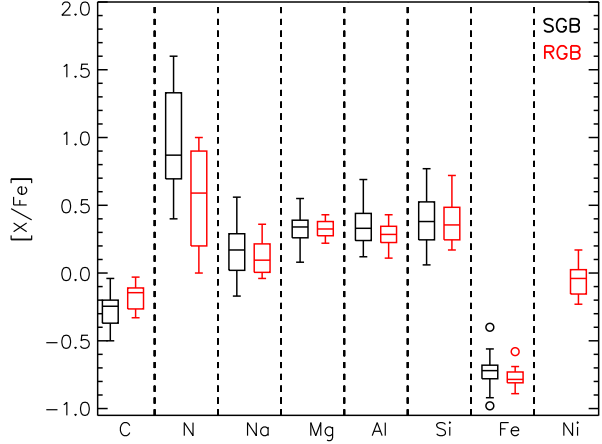}
      \caption{Box and whisker plot of our SGB (black) and RGB (red) abundances. For the non-Fe species, their [X/Fe] relative abundances are shown, for Fe we plotted [Fe/H]. For a given element, the box represents the interquartile range (middle 50\% of the data) and the median is indicated by the horizontal line. The vertical tails extending from the boxes indicate the total range of abundances determined for each element, excluding outliers. Outliers (those 1.5 times the interquartile range) are denoted by open circles.}
        \label{fig:boxGIR}
   \end{figure}

We have inferred abundances for C, N, Na, Al, Si and Mg, all involved in the H-burning at various temperatures, and Ni for the RGB only.
Our elemental abundance results are illustrated in Fig.~\ref{fig:boxGIR}, where we plot separately results for RGB and SGB stars.

No large differences are present in the abundances of the SGB and RGB stars. The most notable differences are observed in N and marginally in C, which are lower and higher, respectively, on the RGB. At a first glance, these offsets appear the opposite than expected if processed material has reached the surface after the first dredge-up at the base of the RGB. On the other hand, these effects are predicted to be higher at lower metallicity, and negligible at the metallicity of 47~Tuc. 
Indeed, our RGB stars are all $\sim$1.5~mag in the $V$ band below the RGB bump, and, except in very metal-poor GCs the evidence of deep-mixing, CN-cycle on the RGB is mainly observed above the bump (e.g., Denissenkov \& VandenBerg\,2003).
We performed a differential abundance study of CH and NH lines using classical 1D hydrostatic and 3D hydrodynamic model stellar atmospheres from the {\sc Stagger} grid (Collet et al.\,2011; Magic et al.\,2013). We verified that 3D effects do not play an important role in generating these differences. Furthermore, these differences diminish after a proper analysis of stellar population is done, as discussed in Sect.~\ref{sec:sgb}.
In the following subsections we will investigate the chemical composition of different photometric sequences by combining results from spectroscopy and photometry.

%
%

\subsection{Multiple stellar populations along the red-giant branch}\label{sec:rgb}

Multiple stellar populations can be easily distinguished from photometry of RGB stars, when CMDs made with $U-B$ and $B-I$ colors are used (e.g., Marino et al.\,2008). The reason is that the $U-B$ color is very sensitive to the abundance of N and C, through the CN and CH molecular bands, while $B-I$ is sensitive to the helium content. Indeed stars with the same luminosity but different helium content have also different effective temperature.
 Recent studies have demonstrated that appropriate combinations of the $U, B, I$ (or the similar {\it HST} filters) provide efficient tools to identify multiple stellar populations in GCs. These include the $U-B$ vs.\,$U-I$ two-color diagram and the $B$ vs.\,$U-B+I$ diagram (Milone et al.\,2012a) and the $B$ vs.\,$C_{\rm U,B,I}=U-2B+I$ diagram introduced by Milone et al.\,(2013) by using {\it HST} filters F336W, F438W (or F410M) and F814W.

\begin{figure}
   \centering
   \includegraphics[width=8.5cm]{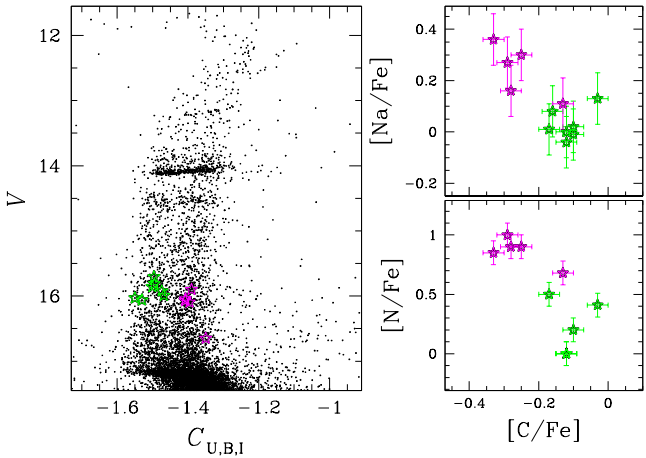}
      \caption{\textit{Left panel:} $V$ vs.\,$C_{\rm U,B,I}$ diagram from ground-based photometry.  \textit{Right panels:} Sodium and Nitrogen abundance relative to iron as a function of [C/Fe]. 
RGB-A and RGB-B target stars ere represented with green and magenta starred symbols, respectively. }
        \label{fig:rgb}
   \end{figure}

Figure~\ref{fig:rgb} shows $V$ vs.\,$C_{\rm U,B,I}$ for stars in 47\,Tuc obtained from the photometric catalog by Stetson (2000), where the two main RGBs and HBs are clearly visible. We have used green and magenta starred symbols to represent the seven RGB-A and the five RGB-B stars studied in this paper.  
 
As shown in the right panels of Fig.~\ref{fig:rgb}, RGB-B stars are enhanced in Na, N, and depleted in C with respect to population-A stars.
Note that, as discussed in Sect.~\ref{CNeffects}, we have assumed the same [O/Fe] for all the stars, which would mostly translate into the RGB-B stars having lower [C/Fe] abundances by $\sim$0.1~dex.
 Confirming previous results, when we consider the entire sample of analyzed RGB stars, C, N, and Na are (anti)-correlated, while no significant variations in Mg have been detected, and the rms for Al is only marginally higher than expected from observational errors.
%
%

\subsection{Multiple stellar populations along the sub-giant branch}\label{sec:sgb}

In Fig.~\ref{fig:CMDhst} we have shown that population-A and population-B stars can  be observed photometrically along all the evolutionary stages, the HB, RGB, SGB, and MS. In contrast, the stellar population associated to the faint SGB has been never identified in any other region of the CMD except for the SGB.

\begin{figure}
   \centering
   \includegraphics[width=8.5cm]{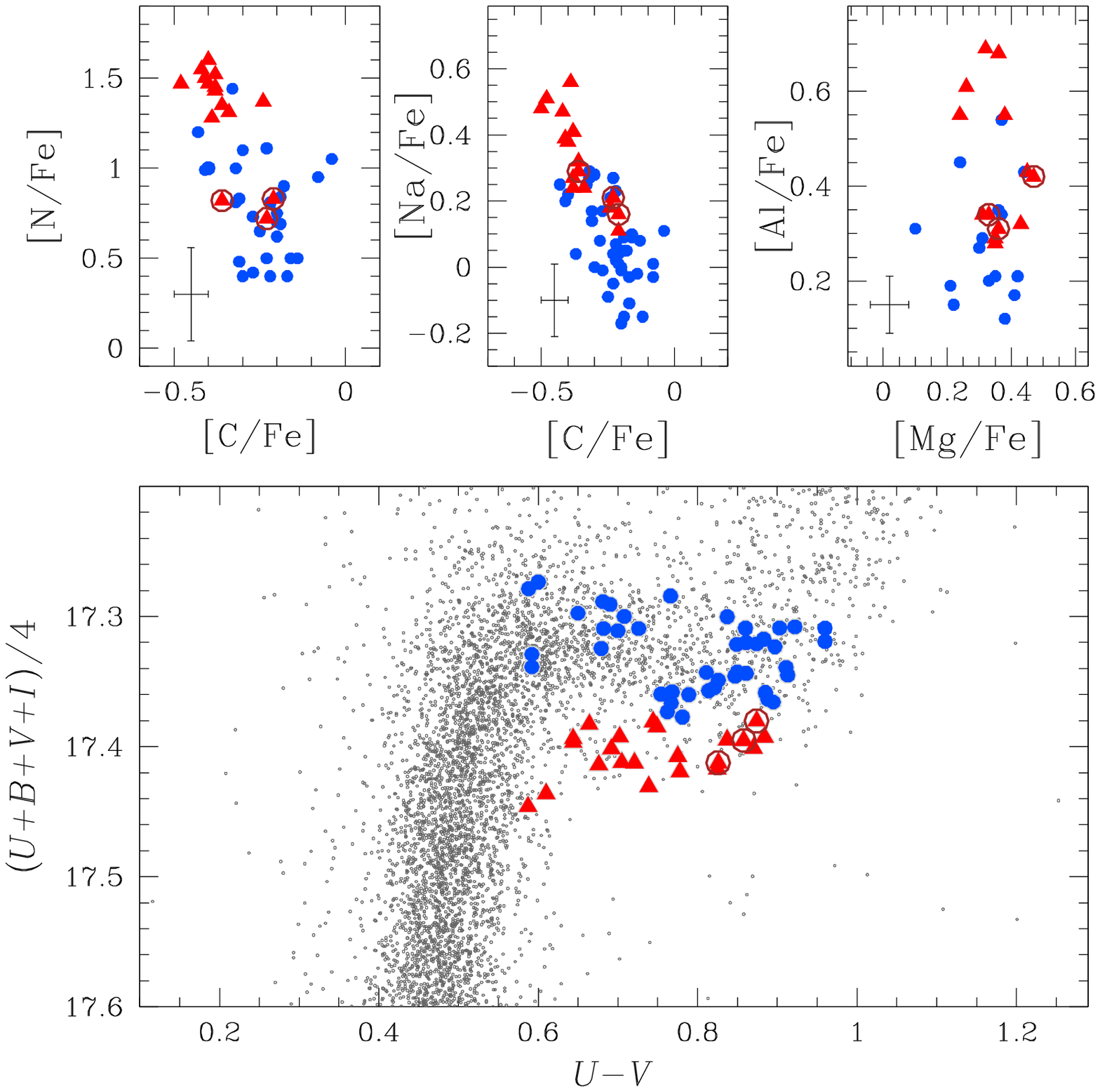}
   \includegraphics[width=8.5cm]{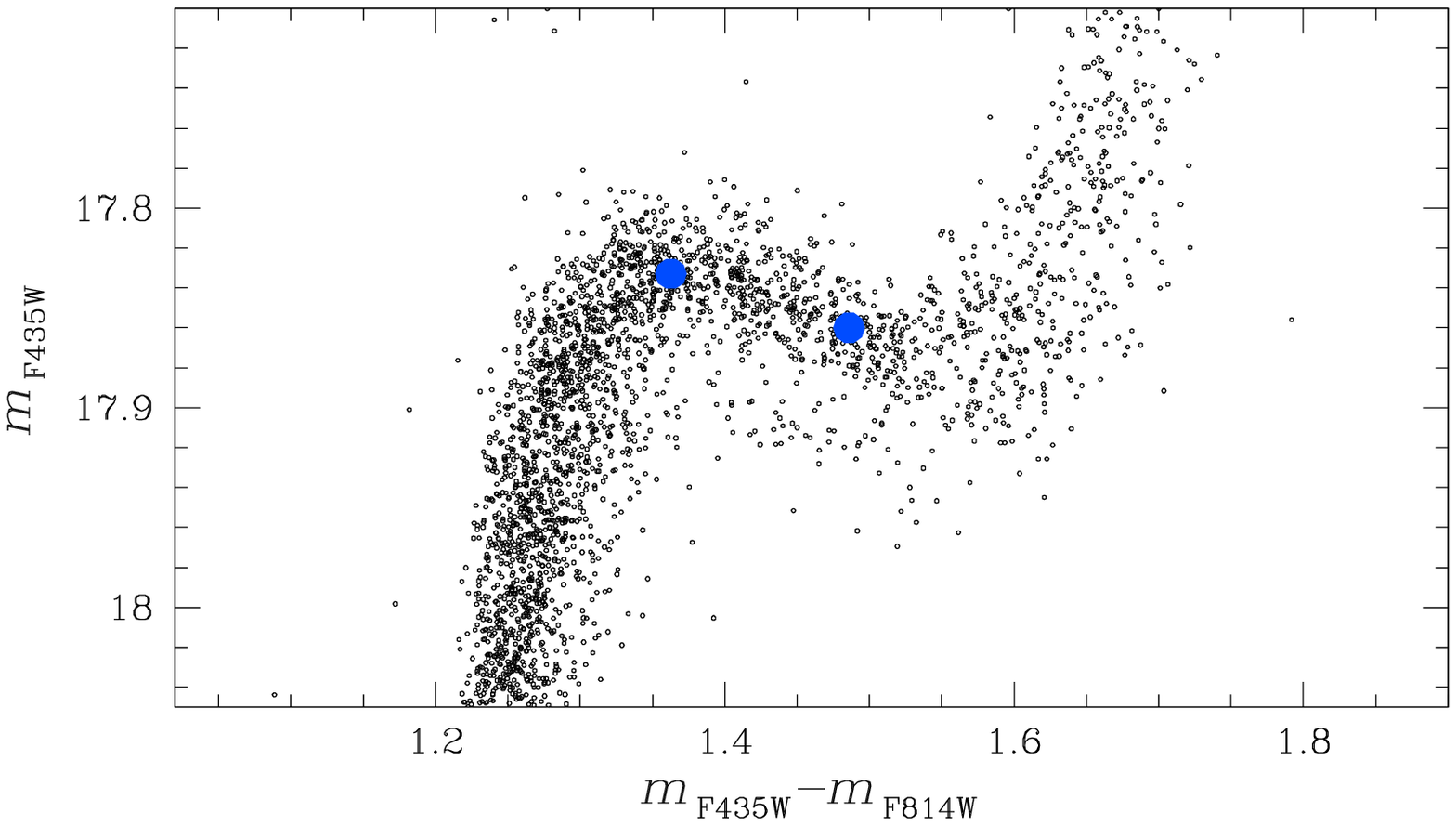}
      \caption{The CMDs plotted in lower panels are obtained from ground-based (middle panel) and {\it HST} photometry (lower panel). The ground-based photometry CMD shows the average of the $U, B, V, I$ magnitudes as a function of the $(U-V)$ color for SGB stars of 47\,Tuc. 
The space-based diagram shows the $m_{F435W}$-$(m_{F435W}-m_{F814W})$ CMD.
Blue dots and red triangles mark the target stars that, on the basis of their position in the CMD, have been classified as bright and faint SGB, respectively. 
The same colors and symbols are used to represent the same stars in the upper paneles where we plot [N/Fe] vs.\,[C/Fe] (left), [Na/Fe] vs.\,[C/Fe] (middle), and [Al/Fe] vs.\,[Mg/Fe] (right).}
        \label{fig:pcap_sgb}
   \end{figure}

By analysing  multi-wavelength {\it HST} photometry of 47\,Tuc, Piotto et al.\,(2012) concluded that, at fixed color, the magnitude separation between the faint SGB and the bulk of SGB stars is almost constant at all the analyzed magnitudes. For this reason, to minimize the magnitude error and highlight the faint SGB, in the middle panel of Fig.~\ref{fig:pcap_sgb} we show the average ($U+B+V+I$)/4 magnitude as a function of the $(U-V)$ color.
In order to explore the chemical composition of different SGB branches, we have selected a sample of stars whose position on the CMD can be more clearly associated with the bright and faint SGB component in the $(U+B+V+I)$/4-$(U-V)$ diagram.
In the lower panel of Fig.~\ref{fig:pcap_sgb} we represent the selected bright and faint SGB stars, with blue dots and red triangles, respectively. 
Two SGB stars fall in the {\it HST} field, and their location on the $m_{F435W}$-$(m_{F435W}-m_{F814W})$ CMD is shown in the lower panel of Fig.~\ref{fig:pcap_sgb}.
These symbols are used consistently throughout the paper. 

Note that in the selection of bright and faint SGBs in Fig.~\ref{fig:pcap_sgb} we have excluded stars with large values of the photometric r.m.s and stars that are not well fitted by the PSF according to their values of chi and sharp. Furthermore, we exclude the possibility that faint SGB stars are blended bright SGBs, as a blend between a SGB star and a MS, SGB, RGB or HB star will result in brighter magnitudes than a single SGB star. The photometric error of the selected bright SGB and faint SGB stars in the adopted $(U+B+V+I)$/4 are on average 0.02~mag and never exceed 0.03~mag. Hence we can safely exclude that most of faint SGB candidates are actually bright SGB stars. Nevertheless,  we cannot exclude some contamination in the selected groups of faint and bright SGBs from the other SGB. In order to address this point we have performed 1,000,000 monte-carlo simulations by randomly adding to each star an error in the $(U-V)$ color and in the $(U+B+V+I)$/4 pseudo-magnitude equal to the observed one. We have found that the contamination of bright SGB stars in the faint SGB sample is smaller than 2, 4, and 5 in the 68.27\%, 95.45\% and 99.73\% of the simulations, respectively.

%
%

%
   \begin{figure}
   \centering
   \includegraphics[width=8.5cm]{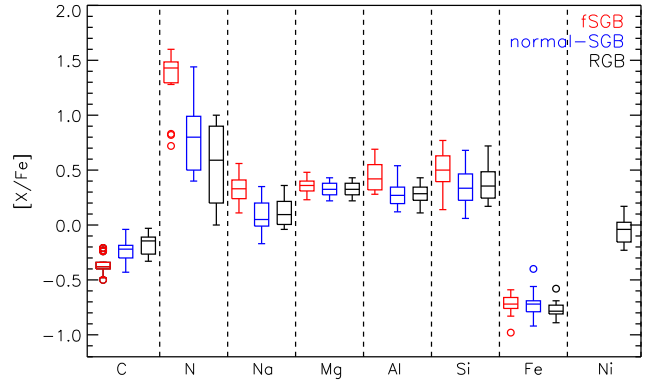}
      \caption{Box and whisker plot of the bright (blue) and faint SGB (red) abundances. The black box plot represents RGB stars. For the non-Fe species, their [X/Fe] relative abundances are shown, for Fe we plotted [Fe/H]. For a given element, the box represents the interquartile range (middle 50\% of the data) and the median is indicated by the horizontal line. The vertical tails extending from the boxes indicate the total range of abundances determined for each element, excluding outliers. Outliers (those 1.5 times the interquartile range) are denoted by open circles.}
        \label{fig:boxGIR1}
   \end{figure}
%

In Fig.~\ref{fig:boxGIR1} we represent the distribution of the inferred chemical abundances for the selected faint and bright SGB stars (red and blue, respectively), with the RGB abundances for comparison (black). Overall, we note that C abundances are on average lower for the faint-SGB, while the bright-SGB average value better agrees with the RGB average abundances. The mean N abundance is higher for faint-SGB stars than bright-SGB ones, while N for the bright-SGB is closer to the value observed on the RGB. This is not surprising, as the progeny of the faint-SGB on the RGB should only account for a tiny fraction of stars. However, although the separation in the two SGB populations weakens the difference in N between the RGB and the bright-SGB stars, the mean abundance of the bright SGB is still higher. This is likely due to biases in our SGB sample with available N abundances. Indeed, due to the weak CN band for SGB stars with low N we were unable to infer abundances for the stars with the lowest N abundances. The presence of this bias is suggested by the mean of the C abundances for the bright SGB stars with missing N abundances [C/Fe]=$-$0.20$\pm$0.03 (rms=0.08), which is approximately the mean abundance of the N poor SGB stars. We note that the rms in [C/Fe] for these SGB stars with unavailable N abundance, 0.08~dex, is relatively small (compared to the values listed in Tab.~\ref{tab:average}), further suggesting that they belong to the same populations with similar N abundances.

In the upper panels of Fig.~\ref{fig:pcap_sgb}, we show the bright- and faint-SGB stars in the [N/Fe]-[C/Fe], [Na/Fe]-[C/Fe] and [Al/Fe]-[Mg/Fe] planes. Bright-SGB stars span a large range in C, N, and Na, being this sequence consistent with internal N-C and Na-C anticorrelations in light elements. 
We note that the estimated internal error in N for SGB stars is quite large, and the observed internal dispersion in the bright-SGB stars only marginally larger. However, the concomitant variation in C is an indication of C-N anticorrelation among bright-SGB stars.   

As represented in the lower panel of Fig.~\ref{fig:pcap_sgb}, the faint SGB and bright SGB populations are distinct sequences, hence their loci in the \teff-\logg\ plane are expected to be different. According to our results, at a given temperature, faint SGB stars have \logg\ higher by a few cents of dex ($<$0.05 dex). At a given surface gravity, the difference in \teff\ is around 50~K, with the faint SGB being warmer. Based on the mean chemical abundance differences listed in Tab.~\ref{tab:average}, only a systematic error of $\sim$300~K on the faint SGB towards lower values can cancel the observed difference in N. 
Our analysis indicates that a large systematic \teff\ error in the direction required to removed the observed chemical differences to be unlikely.

Faint-SGB stars are more clustered around quite high values of N and Na, and lower C abundances. We cannot exclude the possibility of a few faint-SGB stars having higher C, and lower N and Na, as suggested by Fig.~\ref{fig:pcap_sgb}, although we should keep in mind that they can be due to photometric errors causing the contamination of the two selected bright- and faint-SGB samples from stars belonging to the other sequence. 
We do not observe any Al-Mg anticorrelation, and no significant variations in Mg. On the other hand, our results suggest that faint-SGB stars show higher Al abundances than stars distributing on the bright-SGB (right panel).

\begin{figure}
   \centering
   \includegraphics[width=9cm]{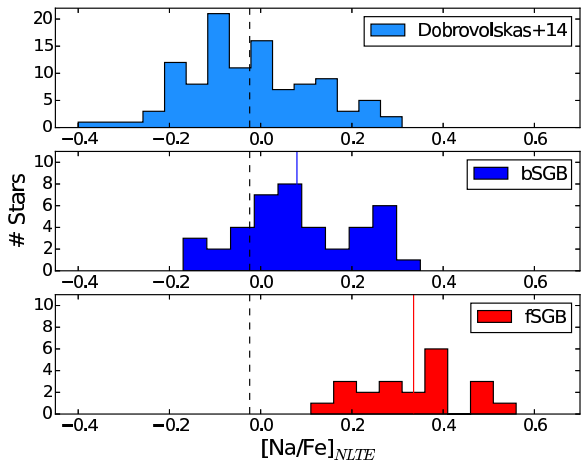}
      \caption{Histogram distributions of the sodium abundance derived by Dobrovolskas et al.\,(2014) for their sample of turn-off stars of 47\,Tucanae (upper panel), and derived in this paper for stars in the bright- (middle panel) and the faint-SGB (lower panel). The black dashed line represents the mean value of [Na/Fe] from Dobrovolskas et al.\,(2014), while the blue (middle panel) and the red (lower panel) solid lines are the means obtained for the two SGB populations in this work.}
        \label{fig:lit_dobro}
   \end{figure}

Dobrovolskas et al.\,(2014, hereafter D14) have used medium-resolution GIRAFFE spectra to determine the sodium abundance for a sample of 110 MS turn-off stars in 47\,Tuc.
In Fig.~\ref{fig:lit_dobro} we provide a comparison of our Na abundances for SGB stars and those from turn-off stars provided by D14. The Na histograms for our stars have been plotted separately for bright and faint-SGB stars. The distribution for the bright-SGB spans a very similar range of that found by D14, with a possible systematic of $\sim$0.10~dex between our data and theirs. On the other hand, the faint-SGB distribution is clearly shifted towards higher abundances, with the presence of high Na values not reached in the D14 sample. 
This is mostly due to the significantly lower fraction of stars corresponding to the population observed in the SGB as a fainter component, which causes a lack/paucity of this population in blind spectroscopic surveys.
Furthermore, the D14 target selection  of turn-off stars might suffer from a ``photometric bias'' towards brighter stars, while the faint SGB turn-off stars lie at higher magnitudes.
We note however that the D14 Na abundance distribution overlaps with the lower half of the faint SGB one, suggesting that a few faint SGB progenitors might be present in their sample.

%
%

\section{Comparison with the literature}

47~Tucanae is one of the most widely spectroscopically studied GC. 
However, most of the studies rely on spectral features present in the optical, so our abundances of many elements have been derived from spectral features that are not commonly used. 
Though results obtained from different lines are subject to systematic differences, it is worth to perform a comparison with some literature work. 
For the comparison with our results, here we chose some of the most recent studies on 47~Tuc or those based on large sample of stars.

\begin{figure*}
   \centering
   \includegraphics[width=13cm]{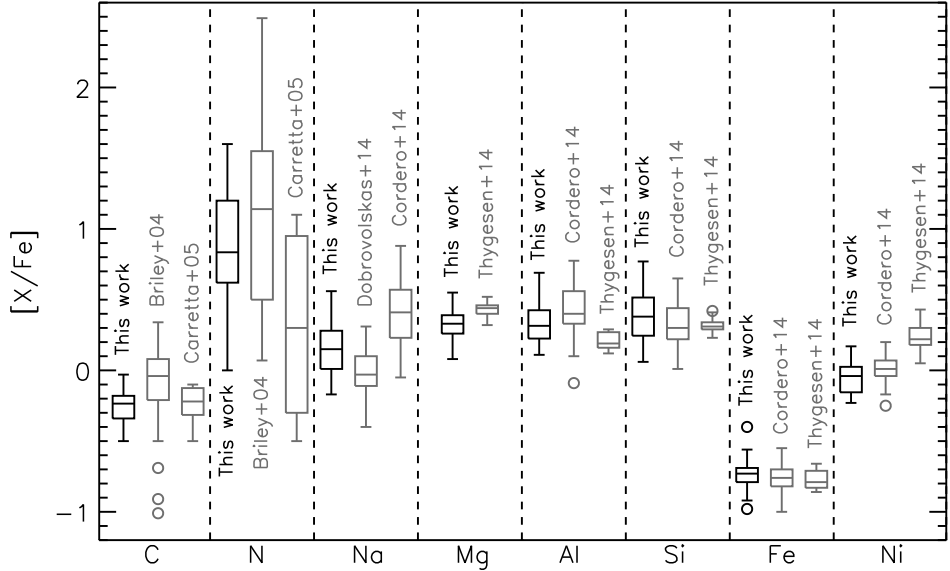}
      \caption{Box and whisker plot of our abundances for the complete sample (black), and from literature studies (gray). For the non-Fe species, their [X/Fe] relative abundances are shown, for Fe we plotted [Fe/H]. For a given element, the box represents the interquartile range (middle 50\% of the data) and the median is indicated by the horizontal line. The vertical tails extending from the boxes indicate the total range of abundances determined for each element, excluding outliers. Outliers (those 1.5 times the interquartile range) are denoted by open circles.}
        \label{fig:boxLIT}
   \end{figure*}

In Fig.~\ref{fig:boxLIT} we compare our abundance distribution with those obtained in previous work. Specifically for C and N we compare with results obtained by Briley et al.\,(2004) based on a large sample of stars mostly on the MS, and Carretta et al.\,(2005) based on 12 TO and SGB stars. Carbon abundances, obtained in all the cases from the CH G band, appear in reasonable agreement, though the Briley et al. sample span a larger range with some stars having lower abundances. Both Briley et al. and Carretta et al. inferred N abundances from the CN UV system at $\sim$3880~\AA. The N abundances from Briley et al. span a larger range than ours, while the Carretta et al. ones are on average lower by $\sim$0.6~dex. 

In Sect.~\ref{sec:sgb} we have compared the Na abundances, corrected for NLTE, for the samples of bright plus faint SGB stars with those from D14, and notice a possible systematic offset of $\sim$0.10~dex among the two sets of data. In Fig.~\ref{fig:boxLIT} we have also plotted the sample of RGBs stars studied by Cordero et al.\,(2014), whose results are based on the doublet $\lambda$6154-6160~\AA. Our results lie in between the Na abundances reported by Dobrovolskas et al.\,(2014) and Cordero et al.\,(2014).

It is worth noticing here, that in Fig.~\ref{fig:boxLIT} we plot our entire sample, which comprises SGBs, including many faint SGB, and RGBs. We expect that our average abundances in those elements like N or Na are higher 
than those obtained from other unbiased samples, given the relative large number of faint SGB stars in our sample.

Our average Mg abundance is lower than that reported by Thygesen et al.\,(2014), being their value [Mg/Fe]=$+$0.43$\pm$0.01 (based on 13 RGB stars), and our value [Mg/Fe]=$+$0.29$\pm$0.01. Part of this difference is probably due to NLTE effects. Indeed, Zhao \& Gehren\,(2000) reported higher NLTE corrections, by $\sim$0.15~dex for a star with \teff/\logg/[A/H]=5780~K/3.50/$-$1.0, for the $\lambda$8213~\AA\ Mg line with respect to the lines in the optical used in most of the studies.

For Al and Si the higher resolution data used in Thygesen et al. gives much smaller abundance ranges. 
However the larger range that we get might be in part due to the bias of our sample towards faint-SGB stars. 
Furthermore, for Si, our uncertainties are much higher ($\sim$0.15~dex) than in previous studies as discussed in Sect.~\ref{CNeffects}.  
Iron abundances agree very well among the compared studies, and finally our Ni abundances are in good agreement with Cordero at al., and lower than in Thygesen et al. 

%
%

\section{The SGB and the total CNO}\label{sec:sgb}

\subsection{Oxygen abundances of the two SGBs}\label{sec:oxygen}

Measurements of individual oxygen abundances from the permitted triplet at $\sim$7770~\AA\ were not possible due to the weakness of these lines and the limited S/N of our spectra. So, individual total C$+$N$+$O abundances could not be inferred, given that we lack information on the actual oxygen abundance of our stars.

It is possible, however, to get an estimate of the average O chemical content for the bright and faint SGBs. 
To this aim, we combined the observed spectra for the stars whose positions on the CMD is more certainly associated with the bright and faint SGB stars, as represented in Fig.~\ref{fig:pcap_sgb}. 
The total number of selected stars in the two sub-samples is 41 and 21, respectively for the bright and faint SGB. Not all of these stars have spectra with sufficient quality to infer individual C and N abundance, but we average all of them with the same weight, so that we obtained two spectra with higher S/N. 

Then, we generated MOOG synthetic spectra for each used star, by assuming the appropriate atmospheric parameters of Sect.~\ref{sec:atm}, and constructed two mean synthetic spectra by averaging them: one for the bright SGB and the other for the faint SGB. Synthetic spectra pairs have been produced for a grid of oxygen abundances from $-$0.4 to 0.8~dex, spaced by 0.1.   
Finally, we computed the spectral synthesis for the observed mean bright and faint SGB spectra by using the constructed grid of synthetic spectra, in the same manner of what was done for the individual spectra (Sect.~\ref{sec:abundances}). 
For the bright SGB we obtain $<$[O/Fe]$>$=$+$0.54, and for the faint SGB $<$[O/Fe]$>$=$+$0.25.

The O triplet is affected by strong NLTE effects. Three-dimensional atmospheric effects impact with minor corrections of a few hundredths of dex (e.g.,  Dobrovolskas et al.\,2014). By considering combined 3D-NLTE effects on the O triplet, in the regime of the atmospheric parameters of our stars, the abundances of O decreases by $\sim$0.15~dex ({\sf inspect-stars.com}; Amarsi et al.\,2016). Hence, we obtain $<$[O/Fe]$>_{3D-NLTE}$=$+$0.39 for the bright SGB, and $<$[O/Fe]$>_{3D-NLTE}$=$+$0.10, for the faint SGB.  

\begin{figure}
   \centering
   \includegraphics[width=8.8cm]{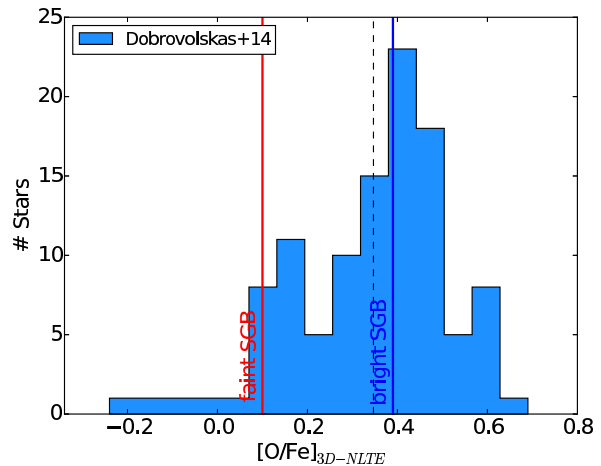}
      \caption{Distribution of [O/Fe] from Dobrovolskas et al.\,(2014). The dashed black line indicates the mean oxygen abundance of this distribution. The blue and red lines are placed on the mean [O/Fe] obtained from the bright and the faint SGB, respectively.}
        \label{fig:O_histo}
   \end{figure}

The comparison between the [O/Fe] distribution obtained for TO stars by D14 in 3D-NLTE and the mean values for the bright and faint SGBs is shown in Fig.~\ref{fig:O_histo}.
Our O mean value for the bright SGB is consistent with the mean [O/Fe] of the stars analysed by D14, which exhibit a O-Na anticorrelation. 
The average O abundance of the faint SGB is much lower. Compared with the distribution obtained by D14 the faint SGB stars have an O slightly lower than the peak of the O-poor/Na-rich stars.
The sample of TO stars studied by D14 does not have any selection on the various photometric sequences, so they mostly belong to the main population corresponding to our bright SGB.
Their data likely include only a few stars corresponding to the minor faint SGB population, possibly including the few stars with low [O/Fe]$<$0. According to our results, this population should have the highest Na abundances.

To validate our procedure used to infer mean O abundances, we applied the same technique to C and N, and found $<$[C/Fe]$>$=$-$0.26/$<$[N/Fe]$>$=$+$0.61 and $<$[C/Fe]$>$=$-$0.38/$<$[N/Fe]$>$=$+$1.20 for the bright and faint SGB, respectively. 
These values are remarkably similar to the mean C and N abundances derived from the mean of the individual abundances obtained for the stars with available abundances: $<$[C/Fe]$>$=$-$0.23$\pm$0.01/$<$[N/Fe]$>$=$+$0.79$\pm$0.05 for the bright SGB, and $<$[C/Fe]$>$=$-$0.36$\pm$0.02/$<$[N/Fe]$>$=$+$1.31$\pm$0.08 for the faint SGB. 

We recall that the averaged spectrum was obtained by using all the stars, even those with missing individual abundances. Specifically, out of 41 bright and 21 faint SGB stars, we have C individual abundances for 40 bright and 18 faint SGBs, and N abundances for only 29 bright SGBs and 15 faint SGBs. 
As the N individual abundances are likely not available for the most N-poor bright SGB stars, we expect that the average value for these stars, $<$[N/Fe]$>$=$+$0.79$\pm$0.05, is over-estimated. Indeed the most relevant discrepancy between the mean abundances from the individual spectra, and the abundances from the averaged spectra, is for the N content of the bright SGB. 
For this reason, we consider the abundances obtained from the averaged spectra as the best estimate of the C and N abundances of the two SGB populations.

\subsection{C$+$N$+$O abundances \label{sec:cno_discussion}}

With the mean abundances of C, N, and O inferred from the averaged spectra, we obtain $\rm {log \epsilon(C+N+O)}$=8.48 and $\rm {log \epsilon(C+N+O)}$=8.54 for the bright SGB and faint SGB, respectively. 
To estimate internal errors for these values, we can assume that the uncertainties associated with the C and N is similar to the errors associated with the average abundances obtained from the individual abundances, e.g. 
$\rm {err_{C,N}}=$0.01,~0.08, and 
$\rm {err_{C,N}}=$0.02,~0.05
for the bright and faint SGB, respectively. As the N individual abundances are likely not tracing all the entire N distribution for the bright SGB, as discussed above, we have increased the error for this population by 0.01, i.e. 
$\rm {err_{N}}$=0.08+0.01. 
Oxygen inferred from the O triplet is sensitive to temperature and surface gravity, and insensitive to metallicity and microturbolence. However, if we consider the uncertainty in the mean \teff\, for the averaged 41 bright and 21 faint SGB stars, assuming that the error associated with each star is 50~K in \teff\, and 0.20~dex in \logg, we get very small errors in these parameters. Taken together these errors affect the mean O abundance by 0.02~dex. The dominant source of error in our O abundances is the limited S/N, which introduces an uncertainty of $\sim$0.08~dex.
The combined internal errors associated to each element, plus the error associated with the solar abundances for C, N and O, which is 0.05~dex (Asplund et al.\,2009), result in internal uncertainties in $\rm {log \epsilon(C+N+O)}$ of 0.07 and 0.08~dex for the bright and the faint SGB, respectively.
Hence, we get $\rm {log \epsilon(C+N+O)}=8.48\pm0.07$ 
for the bright SGB and $\rm {log \epsilon(C+N+O)}=8.54\pm0.08$ for the faint SGB, which means they are consistent within one sigma.

\begin{figure*}
   \centering
   \includegraphics[width=14cm]{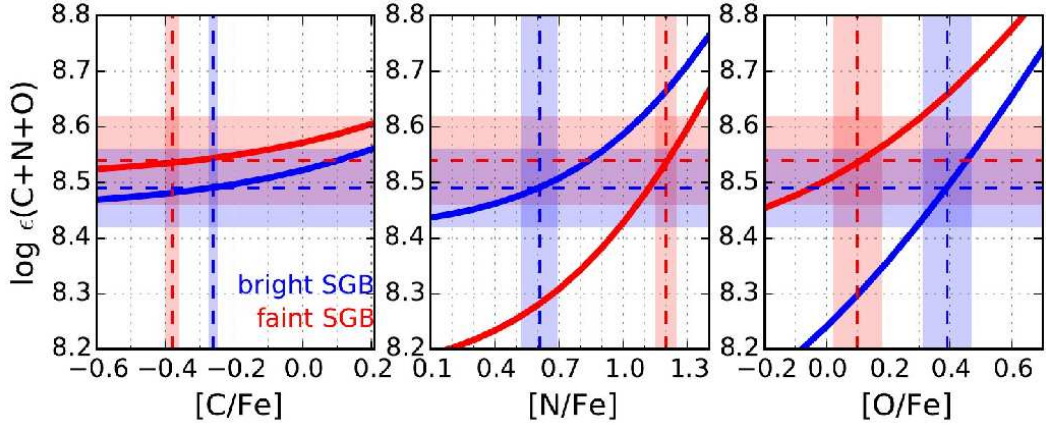}
      \caption{Chemical abundance of the C$+$N$+$O for the bright (solid blue line) and faint SGB (solid red line). From the left to the right the abundance of C, N, and O, have been varied by keeping the other two elements equal to the mean values for the two SGB populations. The dashed lines indicate the obtained total CNO values, and the abundances of C, N, and O inferred from the averaged spectra. Shaded regions represent the estimated internal uncertainties to the various abundances.}
        \label{fig:CNO}
   \end{figure*}

In Fig.~\ref{fig:CNO} we represent the C$+$N$+$O abundance for the bright and faint SGBs as a function of different [C/Fe], [N/Fe], and [O/Fe] values. In each panel, only the abundance of one element has been varied, while the other two elements have been fixed to the abundances obtained from the averaged spectra for the two SGBs. The inferred values of C, N, O, and CNO have been highlighted. Considering internal errors, represented as shaded regions, the two SGBs clearly have different C, N, and O. The CNO value of the faint SGB is slightly higher than that of the bright SGB but our internal errors do not allow us to put firm conclusion on this issue.

Systematic errors in C, N, and O are likely to affect the two SGB populations in similar manner, but not the C$+$N$+$O. Figure~\ref{fig:CNO} illustrates that, at the chemical content of 47~Tuc, C abundances do not impact significantly on the CNO total abundance. If a systematic of $\sim$0.20~dex affects our inferred C abundances, in the range between -0.6 and 0.0, the CNO content will change by no more than $\sim$0.02~dex, and the effect is similar for the two SGBs. Systematics in N and O, that have higher abundances, more significantly affect the CNO abundance of 47~Tuc. A systematic in N by 0.2~dex for the bright SGB (around [N/Fe]$\sim$0.6~dex) will change the CNO by $\sim$0.03~dex, while the CNO of the faint SGB (around [N/Fe]$\sim$1.2~dex) will be affected by $\sim$0.11~dex. Systematics in oxygen, of course, are more important for the bright SGB, with a quite large effect of $\sim$0.15~dex for a variation in [O/Fe] by $\pm$0.2~dex, while the same variation changes the CNO of the faint SGB by $\sim$0.06~dex.
Taken into account possible systematics in C, N or O, the CNO values for the bright and faint SGB will be: 
$\rm {log \epsilon(C+N+O)}=8.48\pm0.07(\pm0.02\pm0.03\pm0.15)$ and 
$\rm {log \epsilon(C+N+O)}=8.54\pm0.08(\pm0.02\pm0.11\pm0.06)$, respectively.
Hence, a positive systematic in N and a negative systematic in O would enhance the CNO difference between the two SGBs, with the fainter being more enhanced in CNO by $\sim$0.15~dex. In this case the difference in CNO would be at a level of 2~$\sigma$. On the other hand, a negative systematic in N or a positive systematic in O would diminish the difference in CNO between the two SGB populations.  
A systematic in O as large as 0.20~dex is unlikely, given that O abundances for RGB from high-resolution spectroscopy and different spectral features give similar values as our ones. We cannot exclude however such large systematics can occur for nitrogen. 

This discussion einlights the range that our CNO abundances might span due to various errors. This is more a qualitative discussion, as we are not able to evaluate the exact systematics and their signs. For instance, different systematics affecting the two SGB populations might also occur, e.g. due to 3D-NLTE effects depending on the abundance and/or to the some uncounted population-effect which could affect our atmospheric parameters derived from photometry. Of course, we can also have combined systematics in C, N, O abundances.

\subsection{The split SGB}
Based on our average C$+$N$+$O values, the faint SGB is enhanced by a factor $\sim$1.1 with respect to the bright SGB, which is within 1~$\sigma$. 
Considering possible systematic errors, the maximum difference we can obtain between the two SGBs is 0.15~dex in log($\epsilon$), which means the faint SGB is enhanced in total C$+$N$+$O by a factor of $\sim$1.4 compared to the bright SGB. 

Theoretical work considering the photometric split on the SGB of 47~Tuc predicts that the two SGBs are populated by stars with different C$+$N$+$O, with the faint SGB being enhanced by a factor 1.4 and diluted by 50\% with pristine material. At the relatively high metallicity of 47~Tuc, the dependence of the SGB location on helium becomes important for the models, and the shape of the faint SGB is consistent with $Y$=0.28 (Di Criscienzo et al.\,2010).

Assuming the faint SGB is enhanced by a factor of 1.1 with respect to the bright SGB, and that the material out of which the faint SGB formed was diluted by the 50\% with pristine material, as interpreted by Di Criscienzo et al., then, the {\it pure} material, prior to any dilution, from which faint SGB stars would have been enhanced in C$+$N$+$O by a factor of 1.3 with respect to the primordial abundance of the bright SGB, consistent with the factor 1.4 predicted by Di Criscienzo et al.  
Considering our C$+$N$+$O average values, and the maximum variation allowed within the systematic errors, which is a factor of 1.4 difference between the two SGBs, the {\it pure} material out of which the faint SGB stars formed would have been enhanced in C$+$N$+$O by a larger factor, 1.8, then diluted by the 50\% with pristine matter.

Furthermore, the higher N and Na, and the lower C and O for the faint SGBs stars is qualitatively consistent with these stars being enhanced in helium by some degree. The isochrones used in Di Criscienzo et al. are consistent with a faint SGB matter being composed of 28\% He by mass.

Di Criscienzo et al.\,(2010) also suggested that the spread within the brighter SGB is likely due to a spread in He mass fraction of 0.02 that can account at the same time for the HB luminosity spread, which is larger than expected from photometric errors. 
Qualitatively, the presence of internal variations in light elements among bright SGB stars is consistent with this picture.

%
%

%
%

\begin{figure*}
   \centering
   \includegraphics[width=9.0cm]{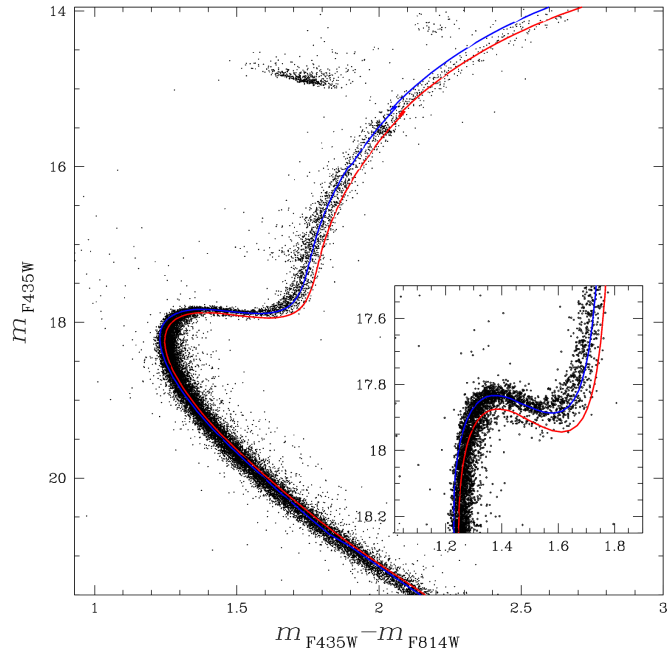}
      \caption{$m_{\rm F435W}$ vs.\,$m_{\rm F435W}-m_{\rm F814W}$ CMD from Milone et al.\,(2012a) with overimposed the CNO-poor (blue line) and CNO-rich (red line) isochrones corresponding to the same average CNO abundances derived for bright- and faint-SGB stars. The inset shows a zoom of the CMD around the SGB.}
        \label{fig:iso}
   \end{figure*}

 To further investigate the role of the CNO abundance on the SGB morphology of 47\,Tuc we have compared the observed CMD with isochrones similar to those of Dotter et al.\,(2008). 
 Specifically, we have generated two sets of isochrones with [Fe/H]=$-$0.73, [$\alpha/Fe$]=0.2 and with the same C$+$N$+$O average abundances inferred for bright SGB and faint SGB stars. We assumed for both isochrones a distance modulus $(m-M)_{0}$=13.29, reddening E$(B-V)$=0.01, primordial helium and age of 13.0~Gyr that provide a good match between the isochrones and the CMD as shown in Fig.~\ref{fig:iso}.
The fact that the magnitude and color difference between the bright and the faint SGB of 47\,Tuc are well reproduced by two coeval isochrones with the same CNO abundances inferred from spectroscopy corroborates the conclusion that the CNO variation is the main factor responsible for the SGB split.
We recall here that the two isochrones have been computed with the same $Y$, while the faint SGB population is likely He-enhanced. To account for this effect would mostly affect the MS and RGB, whose separation in color would be decreased.

%
%

\section{Conclusions}\label{sec:conclusions}

We have presented a spectroscopic analysis of a sample of 74 stars in the GC 47~Tuc, distributed from the SGB up to the base of the RGB. The stars are chosen to lie on both sides of the split SGB in order to investigate their light-element chemical composition.  Following the CMD we found: 
\begin{itemize}
\item{SGB: The major SGB component, the bright SGB, is a {\it normal} component, as observed in all Milky Way GCs, i.e., it is populated by stars with different light elements, and has its own abundance (anti-)correlations. The less-populous, faint SGB component, which accounts for $\sim$10\% of stars in the CMD, is consistent with being composed of stars with higher abundances in the elements produced via hot H-burning, with N and Na abundances being on average higher than the those observed in the bright SGB stars. There is weak evidence for a small C$+$N$+$O enrichment of the faint SGB. The C$+$N$+$O enrichment experienced by this anomalous SGB component is, on average, a factor of $\sim$1.1$\pm$1.3$\pm$0.3 (random plus systematics) over the bright SGB.}
\item{RGB: The two sequences identified by using the $C_{\rm U,B,I}$ index have different content in C, N and Na, and we confirm that the RGB-A and -B are populated by relatively C-rich/N-poor/Na-rich and C-poor/N-rich/Na-poor stars, respectively.}
\end{itemize}

Our results qualitatively agree with theoretical predictions by Di Criscienzo et al.\,(2010) who interpreted the faint SGB as populated by stars enriched in He, with $Y$=0.28, and slightly enriched in C$+$N$+$O by a factor of 1.4 and diluted by 50\% with pristine matter. We do not find any evidence for variations in Fe, as found on the SGBs in M\,22 and NGC\,1851 (Marino et al.\,2012; Gratton et al.\,2012). Given our observational errors, we cannot exclude small enrichments in Fe, by a few hundredths of dex. We note however that 47~Tuc is a quite metal-rich GC, hence if its faint SGB component has experienced the same enrichment as the faint SGBs of M\,22 and NGC\,1851, more pristine Fe-enriched material would have been required to observe the same variations observed in M\,22 and NGC\,1851.

\section*{acknowledgments}
\small
We thank the anonymous referee for his/her helpful suggestions.
AFM, APM and RC have been supported by the Australian Research Council through Discovery Early Career Researcher Awards DE150101816, DE160100851 and DE120102904. C.I.J. gratefully acknowledges support from the Clay Fellowship, administered by the Smithsonian Astrophysical Observatory. 
AD is supported by the Australian Research Council under grant FL110100012.
\normalsize

\bibliographystyle{aa}

\onecolumn
\begin{longtable}{lccccccccrr}
\caption{Coordinates, basic photometric data and radial velocities for the stars in the field of view of 47~Tucanae. The listed magnitudes are corrected for differential reddening.}\label{tab:data}\\ \hline\hline
ID            &  RA (J2000) & DEC (J2000) & $B$ & $V$ & RV [\kmsec]          \\
\hline
\endfirsthead
\caption{continued.}\\\hline\hline
ID            &  RA (J2000) & DEC (J2000) & $B$ & $V$ & RV [\kmsec]          \\
\hline                                                                   
\endhead
\endfoot
 N104e-51475  & 00:24:01.80 & $-$71:59:58.2 & 17.915 & 17.267 &  $-$15.90  \\
 N104e-31450  & 00:24:00.23 & $-$72:01:07.5 & 16.658 & 15.811 &  $-$27.28  \\
  N104e-4888  & 00:23:21.40 & $-$72:00:21.8 & 16.913 & 16.094 &  $-$24.53  \\     
 N104e-21122  & 00:23:36.22 & $-$72:01:54.2 & 17.805 & 17.172 &  $-$28.20  \\    
 N104e-11207  & 00:23:42.16 & $-$72:01:20.1 & 17.851 & 17.122 &  $-$10.74  \\    
 N104e-51104  & 00:23:34.59 & $-$71:57:13.2 & 17.928 & 17.255 &  $-$19.24  \\ 	      
  N104e-4647  & 00:23:04.49 & $-$71:57:34.1 & 16.869 & 16.042 &  $-$11.53  \\     
  N104e-1610  & 00:23:01.41 & $-$71:56:35.9 & 17.931 & 17.176 &  $-$13.65  \\     
  N104e-1716  & 00:23:10.13 & $-$71:59:45.7 & 17.869 & 17.134 &  $-$10.44  \\     
 N104e-11887  & 00:24:30.21 & $-$71:55:36.4 & 17.911 & 17.196 &  $-$24.72  \\    
 N104e-11495  & 00:24:03.08 & $-$72:02:41.6 & 17.856 & 17.156 &  $-$25.53  \\    
 N104e-41810  & 00:24:25.55 & $-$71:58:17.3 & 16.818 & 15.989 &  $-$26.21  \\ 	      
 N104i-200149 & 00:24:20.13 & $-$72:04:05.5 & 17.425 & 16.648 &   $-$6.47  \\  
 N104e-31517  & 00:24:04.32 & $-$71:58:33.4 & 16.860 & 16.030 &  $-$23.23  \\      
 N104e-11211  & 00:23:42.44 & $-$71:59:44.7 & 17.892 & 17.207 &  $-$10.60  \\    
 N104e-42244  & 00:24:50.95 & $-$71:59:45.5 & 16.874 & 16.071 &  $-$23.40  \\    
 N104e-52565  & 00:25:12.42 & $-$71:56:25.6 & 17.890 & 17.272 &  $-$19.00  \\    
 N104e-52434  & 00:25:02.43 & $-$71:56:17.2 & 17.928 & 17.208 &  $-$29.52  \\   
 N104e-12252  & 00:24:51.09 & $-$71:57:39.4 & 17.887 & 17.147 &  $-$14.54  \\    
 N104e-12344  & 00:24:56.73 & $-$71:53:36.6 & 17.869 & 17.101 &  $-$23.57  \\    
 N104e-32006  & 00:24:37.19 & $-$71:57:45.1 & 16.692 & 15.856 &  $-$26.42  \\    
 N104e-52518  & 00:25:10.24 & $-$72:03:40.8 & 17.877 & 17.253 &  $-$13.35  \\   
 N104e-52196  & 00:24:48.20 & $-$72:03:38.1 & 17.926 & 17.277 &   $-$9.21  \\    
 N104e-52589  & 00:25:14.19 & $-$72:01:15.9 & 17.955 & 17.240 &  $-$18.56  \\    
 N104e-12303  & 00:24:55.25 & $-$72:01:13.8 & 17.919 & 17.227 &  $-$13.45  \\    
 N104e-32271  & 00:24:52.69 & $-$72:00:52.7 & 16.560 & 15.706 &   $-$7.45  \\     
 N104e-52066  & 00:24:41.10 & $-$72:01:24.3 & 17.916 & 17.217 &  $-$21.55  \\    
 N104e-32766  & 00:25:28.86 & $-$72:03:19.8 & 16.888 & 16.062 &  $-$14.85  \\    
 N104e-12977  & 00:25:45.44 & $-$72:05:10.8 & 17.864 & 17.166 &  $-$18.73  \\    
 N104e-53309  & 00:26:19.82 & $-$72:04:16.9 & 17.956 & 17.382 &  $-$14.26  \\   
 N104e-42442  & 00:25:04.12 & $-$72:04:32.4 & 16.717 & 15.900 &  $-$25.17  \\      
 N104e-13038  & 00:25:49.98 & $-$72:02:38.2 & 17.918 & 17.176 &  $-$16.31  \\    
 N104e-53193  & 00:26:04.04 & $-$72:02:01.4 & 17.921 & 17.268 &  $-$20.16  \\   
 N104e-22676  & 00:25:22.31 & $-$72:07:26.3 & 17.816 & 17.172 &  $-$15.40  \\    
 N104e-23387  & 00:26:36.15 & $-$72:07:43.2 & 17.857 & 17.213 &  $-$20.20  \\    
 N104e-12681  & 00:25:22.34 & $-$72:05:24.7 & 17.835 & 17.135 &  $-$20.67  \\    
 N104e-52912  & 00:25:41.95 & $-$72:07:05.1 & 17.904 & 17.251 &  $-$10.96  \\   
 N104e-52849  & 00:25:37.34 & $-$72:06:14.7 & 17.914 & 17.230 &  $-$22.55  \\    
 N104e-32404  & 00:25:01.75 & $-$72:05:33.4 & 16.768 & 15.924 &  $-$21.11  \\    
 N104e-52324  & 00:24:57.63 & $-$72:11:49.7 & 17.918 & 17.327 &  $-$17.52  \\    
 N104e-22622  & 00:25:18.29 & $-$72:08:52.0 & 17.780 & 17.170 &  $-$20.26  \\      
 N104e-12978  & 00:25:46.64 & $-$72:12:18.5 & 17.888 & 17.172 &  $-$18.91  \\    
 N104e-52983  & 00:25:46.68 & $-$72:09:38.0 & 17.927 & 17.215 &   $-$9.20  \\   
 N104e-52203  & 00:24:49.32 & $-$72:09:24.1 & 17.896 & 17.251 &   $-$7.08  \\    
 N104e-12511  & 00:25:10.52 & $-$72:12:08.5 & 17.826 & 17.152 &  $-$19.12  \\    
 N104e-51107  & 00:23:35.37 & $-$72:08:34.6 & 17.887 & 17.246 &  $-$11.22  \\    
 N104e-12576  & 00:25:15.00 & $-$72:13:36.4 & 17.886 & 17.144 &  $-$23.05  \\    
 N104e-11195  & 00:23:41.62 & $-$72:11:38.6 & 17.876 & 17.169 &  $-$32.06  \\    
 N104e-11459  & 00:24:01.46 & $-$72:09:34.3 & 17.874 & 17.115 &  $-$12.28  \\    
 N104e-11473  & 00:24:02.37 & $-$72:10:55.3 & 17.838 & 17.127 &  $-$18.76  \\    
 N104e-21389  & 00:23:56.57 & $-$72:07:11.9 & 17.914 & 17.327 &   $-$9.20  \\    
 N104e-11602  & 00:24:11.54 & $-$72:13:37.3 & 17.924 & 17.170 &  $-$19.24  \\     
 N104e-11897  & 00:24:32.01 & $-$72:12:56.0 & 17.864 & 17.164 &  $-$16.50  \\    
  N104e-1828  & 00:23:17.95 & $-$72:08:02.4 & 17.856 & 17.134 &   $-$4.26  \\     
 N104e-11004  & 00:23:28.34 & $-$72:11:49.7 & 17.914 & 17.201 &  $-$10.08  \\    
 N104e-11197  & 00:23:41.68 & $-$72:10:53.8 & 17.823 & 17.125 &  $-$16.59  \\    
  N104e-1626  & 00:23:03.00 & $-$72:03:43.6 & 17.882 & 17.196 &   $-$4.29  \\     
 N104e-11119  & 00:23:36.36 & $-$72:10:32.0 & 17.891 & 17.173 &  $-$24.60  \\    
 N104i-201037 & 00:23:48.81 & $-$72:05:40.6 & 17.811 & 17.112 &   $-$5.29  \\  
 N104e-11040  & 00:23:30.28 & $-$72:09:52.6 & 17.861 & 17.141 &  $-$20.89  \\    
  N104e-2799  & 00:23:16.18 & $-$72:07:34.7 & 17.805 & 17.216 &   $-$9.07  \\  	      
  N104e-1765  & 00:23:13.59 & $-$72:10:44.0 & 17.858 & 17.189 &  $-$14.82  \\     
  N104e-5249  & 00:22:22.46 & $-$72:06:42.5 & 17.899 & 17.256 &   $-$7.82  \\    
  N104e-5383  & 00:22:36.97 & $-$72:06:40.2 & 17.891 & 17.276 &  $-$20.20  \\    
  N104e-1544  & 00:22:55.04 & $-$72:07:53.5 & 17.892 & 17.173 &  $-$12.02  \\     
  N104e-2739  & 00:23:11.86 & $-$72:07:31.5 & 17.769 & 17.162 &  $-$13.44  \\     
  N104e-1326  & 00:22:30.86 & $-$72:10:09.2 & 17.845 & 17.153 &  $-$27.96  \\  	      
  N104e-1509  & 00:22:51.54 & $-$72:10:29.8 & 17.863 & 17.109 &  $-$12.41  \\     
  N104e-1781  & 00:23:14.72 & $-$72:03:32.4 & 17.873 & 17.127 &  $-$16.15  \\     
  N104e-1830  & 00:23:18.04 & $-$72:07:07.5 & 17.868 & 17.183 &   $-$4.95  \\     
  N104e-1148  & 00:22:08.79 & $-$72:04:31.3 & 17.856 & 17.169 &  $-$21.53  \\       
  N104e-5134  & 00:22:04.08 & $-$72:05:50.0 & 17.890 & 17.212 &  $-$12.52  \\      
  N104e-5623  & 00:23:02.50 & $-$72:02:04.7 & 17.879 & 17.230 &  $-$11.32  \\     
 N104e-11060  & 00:23:31.33 & $-$72:03:11.4 & 17.893 & 17.190 &  $-$18.29  \\     
  N104e-2717  & 00:23:10.22 & $-$72:00:59.1 & 17.835 & 17.225 &  $-$13.24  \\     
\hline
\end{longtable}

\onecolumn
\begin{longtable}{l cccc c c c r c c c r}
\caption{Adopted atmospheric parameters and inferred chemical abundances of 47~Tucanae.}\label{tab:abb} \\ \hline\hline
ID & \teff\ & \logg & \vmicro & [Fe/H] & [C/Fe] & [N/Fe] & [Na/Fe]$_{\rm {LTE}}$ & [Na/Fe]$_{\rm {NLTE}}$ & [Mg/Fe] & [Al/Fe] & [Si/Fe] & [Ni/Fe] \\ 
\hline
\endfirsthead
\caption{continued.}\\\hline\hline
ID & \teff\ & \logg & \vmicro & [Fe/H] & [C/Fe] & [N/Fe] & [Na/Fe]$_{\rm {LTE}}$ & [Na/Fe]$_{\rm {NLTE}}$ & [Mg/Fe] & [Al/Fe] & [Si/Fe] & [Ni/Fe] \\ 
\hline
\endhead
\endfoot
 N104e-51475 & 5621 & 3.90 & 1.07 & $-$0.72 & $-$0.38 &   1.52 & 0.78 &    0.41 &   0.35 &  0.28 & 0.67 &    --    \\ 
 N104e-31450 & 5234 & 3.17 & 1.22 & $-$0.58 & $-$0.03 &   0.41 & 0.55 &    0.13 &   0.32 &  0.24 & 0.39 &    0.17  \\ 
  N104e-4888 & 5059 & 3.20 & 1.18 & $-$0.78 & $-$0.13 &   0.68 & 0.49 &    0.11 &   0.33 &  0.21 & 0.45 & $-$0.04  \\ 
 N104e-21122 & 5670 & 3.88 & 1.08 & $-$0.65 & $-$0.24 &    --  & 0.61 &    0.21 &   0.24 &  0.45 & 0.15 &    --    \\ 
 N104e-11207 & 5367 & 3.74 & 1.02 & $-$0.70 & $-$0.17 &   0.40 & 0.35 & $-$0.03 &   0.41 &  0.17 & 0.37 &    --    \\ 
 N104e-51104 & 5540 & 3.86 & 1.05 & $-$0.71 & $-$0.39 &   1.28 & 0.88 &    0.56 &   0.43 &  0.32 & 0.77 &    --    \\ 
  N104e-4647 & 5064 & 3.19 & 1.18 & $-$0.72 & $-$0.28 &   0.90 & 0.53 &    0.16 &   0.43 &  0.31 & 0.50 & $-$0.19  \\ 
  N104e-1610 & 5290 & 3.73 & 1.01 & $-$0.71 & $-$0.23 &   0.50 & 0.32 & $-$0.05 &   0.31 &    -- & 0.54 &    --    \\ 
  N104e-1716 & 5349 & 3.74 & 1.02 & $-$0.72 & $-$0.21 &    --  & 0.38 &    0.01 &   0.17 &    -- & 0.21 &    --    \\ 
 N104e-11887 & 5409 & 3.79 & 1.03 & $-$0.56 & $-$0.20 &   0.70 & 0.37 & $-$0.01 &   0.36 &    -- & 0.68 &    --    \\ 
 N104e-11495 & 5455 & 3.79 & 1.04 & $-$0.82 &   --    &    --  & 0.70 &    0.35 &   0.35 &  0.21 & 0.33 &    --    \\ 
 N104e-41810 & 4920 & 3.10 & 1.18 & $-$0.84 & $-$0.17 &   0.50 & 0.38 &    0.01 &   0.27 &  0.33 & 0.36 & $-$0.04  \\ 
N104i-200149 & 5216 & 3.50 & 1.22 & $-$0.89 & $-$0.33 &   0.85 & 0.71 &    0.36 &   0.22 & 0.36  & 0.23 &    --    \\
 N104e-31517 & 5102 & 3.20 & 1.19 & $-$0.78 & $-$0.10 &    --  & 0.42 &    0.02 &   0.32 &  0.40 & 0.34 & $-$0.03  \\ 
 N104e-11211 & 5502 & 3.83 & 1.04 & $-$0.66 & $-$0.18 &   0.90 & 0.44 &    0.05 &   0.24 &   --  & 0.34 &    --    \\ 
 N104e-42244 & 5141 & 3.23 & 1.19 & $-$0.69 & $-$0.29 &   1.00 & 0.64 &    0.27 &   0.40 &  0.43 & 0.49 &    0.08  \\ 
 N104e-52565 & 5721 & 3.93 & 1.09 & $-$0.68 & $-$0.38 &    --  & 0.67 &    0.27 &   0.41 &  --   & 0.64 &    --    \\ 
 N104e-52434 & 5394 & 3.79 & 1.02 & $-$0.73 & $-$0.36 &   0.82 & 0.64 &    0.29 &   0.33 &  0.34 & 0.50 &    --    \\ 
 N104e-12252 & 5334 & 3.74 & 1.02 & $-$0.71 & $-$0.31 &   0.83 & 0.52 &    0.17 &   0.41 &  --   & 0.65 &    --    \\ 
 N104e-12344 & 5253 & 3.69 & 1.01 & $-$0.73 & $-$0.37 &    --  & 0.40 &    0.04 &   0.34 &  --   & --   &    --    \\ 
 N104e-32006 & 5075 & 3.12 & 1.19 & $-$0.79 & $-$0.12 &   0.00 & 0.41 &    0.00 &   0.36 &  0.32 & 0.26 & $-$0.20  \\ 
 N104e-52518 & 5701 & 3.92 & 1.09 & $-$0.78 & $-$0.38 &   1.43 & 0.64 &    0.24 &   0.26 &  0.61 & 0.46 &    --    \\ 
 N104e-52196 & 5617 & 3.90 & 1.06 & $-$0.60 & $-$0.38 &   1.45 & 0.77 &    0.41 &   0.36 &  0.68 & 0.57 &    --    \\ 
 N104e-52589 & 5409 & 3.81 & 1.02 & $-$0.59 & $-$0.23 &   0.72 & 0.57 &    0.21 &   0.47 &  0.42 & 0.50 &    --    \\ 
 N104e-12303 & 5480 & 3.83 & 1.04 & $-$0.61 &     --  &    --  & 0.68 &    0.33 &   0.32 &  0.69 & 0.62 &    --    \\ 
 N104e-32271 & 4981 & 3.01 & 1.18 & $-$0.80 & $-$0.10 &   0.20 & 0.39 & $-$0.01 &   0.33 &  0.26 & 0.42 & $-$0.12  \\ 
 N104e-52066 & 5458 & 3.82 & 1.03 & $-$0.66 & $-$0.42 &   1.55 & 0.80 &    0.47 &   0.48 &  --   & 0.35 &    --    \\ 
 N104e-32766 & 5076 & 3.20 & 1.18 & $-$0.79 & $-$0.12 &   0.00 & 0.36 & $-$0.04 &   0.28 &  0.14 & 0.33 &    0.12  \\ 
 N104e-12977 & 5461 & 3.80 & 1.04 & $-$0.75 & $-$0.22 &   0.81 & 0.40 &    0.02 &   0.34 &  --   & 0.35 &    --    \\ 
 N104e-42442 & 5098 & 3.14 & 1.19 & $-$0.82 & $-$0.25 &   0.90 & 0.66 &    0.30 &   0.25 &  0.26 & 0.36 & $-$0.23  \\ 
 N104e-13038 & 5328 & 3.75 & 1.02 & $-$0.70 & $-$0.30 &   0.40 & 0.37 &    0.00 &   0.33 &  0.20 & 0.62 &    --    \\ 
 N104e-53193 & 5604 & 3.89 & 1.06 & $-$0.69 & $-$0.36 &   1.35 & 0.69 &    0.32 &   0.38 &  --   & 0.14 &    --    \\ 
 N104e-22676 & 5634 & 3.86 & 1.07 & $-$0.92 & $-$0.12 &    --  & 0.27 & $-$0.15 &   0.18 &  --   & --   &    --    \\ 
 N104e-23387 & 5634 & 3.88 & 1.07 & $-$0.40 & $-$0.32 &   1.00 & 0.67 &    0.29 &   0.55 &  --   & --   &    --    \\ 
 N104e-12681 & 5455 & 3.78 & 1.04 & $-$0.76 & $-$0.43 &   1.20 & 0.60 &    0.25 &   0.39 &  --   & --   &    --    \\ 
 N104e-52912 & 5604 & 3.89 & 1.06 & $-$0.63 & $-$0.50 &    --  & 0.83 &    0.48 &   0.38 &  --   & 0.47 &    --    \\ 
 N104e-52849 & 5505 & 3.84 & 1.04 & $-$0.75 & $-$0.21 &    --  & 0.49 &    0.11 &   0.40 &  --   & 0.65 &    --    \\ 
 N104e-32404 & 5069 & 3.14 & 1.19 & $-$0.74 & $-$0.16 &    --  & 0.47 &    0.08 &   0.43 &  0.11 & 0.30 & $-$0.05  \\ 
 N104e-52324 & 5814 & 3.99 & 1.11 & $-$0.72 & $-$0.41 &   1.50 & 0.77 &    0.39 &   0.30 &  --   & 0.29 &    --    \\ 
 N104e-22622 & 5748 & 3.90 & 1.10 & $-$0.71 & $-$0.33 &   1.44 & 0.65 &    0.25 &   0.27 &  --   & 0.24 &    --    \\ 
 N104e-12978 & 5406 & 3.78 & 1.03 & $-$0.66 & $-$0.23 &    --  & 0.42 &    0.04 &   0.47 &  --   & 0.44 &    --    \\ 
 N104e-52983 & 5418 & 3.80 & 1.03 & $-$0.69 & $-$0.21 &    --  & 0.42 &    0.04 &   0.42 &  0.21 & 0.38 &    --    \\ 
 N104e-52203 & 5631 & 3.89 & 1.07 & $-$0.79 & $-$0.34 &   1.31 & 0.62 &    0.24 &   0.35 &  0.29 & 0.44 &    --    \\ 
 N104e-12511 & 5537 & 3.82 & 1.05 & $-$0.84 & $-$0.19 &    --  & 0.26 & $-$0.15 &   0.23 &  --   & --   &    --    \\ 
 N104e-51107 & 5644 & 3.90 & 1.07 & $-$0.79 & $-$0.48 &   1.47 & 0.86 &    0.51 &   0.31 &  0.34 & 0.49 &    --    \\ 
 N104e-12576 & 5328 & 3.74 & 1.02 & $-$0.79 & $-$0.31 &   0.48 & 0.50 &    0.14 &   0.36 &  --   & 0.32 &    --    \\ 
 N104e-11195 & 5434 & 3.79 & 1.03 & $-$0.74 & $-$0.27 &   0.73 & 0.54 &    0.17 &   0.26 &  --   & 0.17 &    --    \\ 
 N104e-11459 & 5279 & 3.70 & 1.01 & $-$0.72 & $-$0.28 &    --  & 0.44 &    0.08 &   0.40 &  --   & 0.41 &    --    \\ 
 N104e-11473 & 5421 & 3.77 & 1.03 & $-$0.76 & $-$0.32 &   0.81 & 0.63 &    0.27 &   0.37 &  0.34 & 0.45 &    --    \\ 
 N104e-21389 & 5829 & 3.99 & 1.12 & $-$0.98 &     --  &    --  & 0.79 &    0.40 &   0.26 &  --   & --   &    --    \\ 
 N104e-11602 & 5293 & 3.73 & 1.01 & $-$0.76 & $-$0.27 &   0.42 & 0.35 & $-$0.01 &   0.26 &  --   & --   &    --    \\ 
 N104e-11897 & 5455 & 3.79 & 1.04 & $-$0.87 & $-$0.19 &   0.69 & 0.47 &    0.09 &   0.08 &  --   & --   &    --    \\ 
  N104e-1828 & 5388 & 3.76 & 1.03 & $-$0.71 & $-$0.22 &   0.40 & 0.44 &    0.07 &   0.26 &  --   & 0.17 &    --    \\ 
 N104e-11004 & 5415 & 3.79 & 1.03 & $-$0.76 &   --    &    --  & 0.65 &    0.30 &   0.24 &  0.55 & 0.72 &    --    \\ 
 N104e-11197 & 5461 & 3.78 & 1.04 & $-$0.79 & $-$0.17 &    --  & 0.29 & $-$0.11 &   0.25 &  --   & --   &    --    \\ 
  N104e-1626 & 5499 & 3.82 & 1.04 & $-$0.63 & $-$0.19 &   0.84 & 0.44 &    0.05 &   0.40 &  --   & 0.49 &    --    \\ 
 N104e-11119 & 5400 & 3.78 & 1.03 & $-$0.62 & $-$0.20 &   0.62 & 0.38 &    0.00 &   0.38 &  0.12 & 0.22 &    --    \\ 
N104i-201037 & 5488 & 3.79 & 1.05 & $-$0.82 & $-$0.22 &   0.80 & 0.59 &    0.23 &   0.10 &  0.31 & 0.06 &    --    \\ 
 N104e-11040 & 5394 & 3.76 & 1.03 & $-$0.70 & $-$0.41 &   0.99 & 0.56 &    0.20 &   0.36 &  0.35 & 0.55 &    --    \\ 
  N104e-2799 & 5821 & 3.94 & 1.12 & $-$0.80 & $-$0.04 &   1.05 & 0.53 &    0.11 &   0.23 &  --   & 0.12 &    --    \\ 
  N104e-1765 & 5553 & 3.84 & 1.06 & $-$0.80 & $-$0.08 &   0.95 & 0.37 & $-$0.03 &   0.21 &  0.19 & 0.33 &    --    \\ 
  N104e-5249 & 5637 & 3.90 & 1.07 & $-$0.83 & $-$0.24 &   1.37 & 0.57 &    0.18 &   0.38 &  0.55 & --   &    --    \\ 
  N104e-5383 & 5731 & 3.94 & 1.09 & $-$0.73 & $-$0.40 &   1.60 & 0.77 &    0.38 &   0.23 &  --   & 0.23 &    --    \\ 
  N104e-1544 & 5397 & 3.77 & 1.03 & $-$0.69 & $-$0.16 &   0.50 & 0.47 &    0.09 &   0.30 &  0.27 & 0.32 &    --    \\ 
  N104e-2739 & 5759 & 3.90 & 1.11 & $-$0.87 & $-$0.08 &    --  & 0.44 &    0.01 &   0.22 &  --   & 0.25 &    --    \\ 
  N104e-1326 & 5480 & 3.80 & 1.04 & $-$0.77 & $-$0.25 &   0.65 & 0.31 & $-$0.09 &   0.42 &  --   & --   &    --    \\ 
  N104e-1509 & 5293 & 3.71 & 1.01 & $-$0.71 & $-$0.40 &   1.00 & 0.56 &    0.22 &   0.44 &  0.43 & 0.48 &    --    \\ 
  N104e-1781 & 5317 & 3.72 & 1.02 & $-$0.83 & $-$0.20 &   0.75 & 0.21 & $-$0.17 &   0.22 &  0.15 & 0.16 &    --    \\ 
  N104e-1830 & 5502 & 3.82 & 1.04 & $-$0.62 & $-$0.23 &   1.11 & 0.63 &    0.27 &   0.28 &  --   & 0.51 &    --    \\ 
  N104e-1148 & 5496 & 3.81 & 1.04 & $-$0.75 & $-$0.14 &   0.50 & 0.37 & $-$0.02 &   0.31 &  0.29 & 0.33 &    --    \\ 
  N104e-5134 & 5524 & 3.84 & 1.05 & $-$0.63 & $-$0.30 &   1.10 & 0.65 &    0.28 &   0.37 &  0.54 & 0.38 &    --    \\ 
  N104e-5623 & 5617 & 3.88 & 1.07 & $-$0.59 & $-$0.40 &   1.47 & 0.74 &    0.38 &   0.45 &  0.43 & 0.54 &    --    \\ 
 N104e-11060 & 5446 & 3.80 & 1.03 & $-$0.73 & $-$0.21 &   0.83 & 0.52 &    0.16 &   0.36 &  0.31 & 0.14 &    --    \\ 
  N104e-2717 & 5748 & 3.92 & 1.10 & $-$0.78 & $-$0.13 &    --  & 0.50 &    0.08 &   0.33 &  --   & 0.23 &    --    \\ 
\hline
\end{longtable}

\begin{table*}
\caption{Sensitivity of derived GIRAFFE abundances to the uncertainties in atmospheric parameters, the limilted S/N ($\sigma_{\rm fit}$) and the total error due to these contributions ($\sigma_{\rm tot}$). Two entries correspond to different contributions for SGB/RGB stars.\label{tab:errGIR}}
\begin{tabular}{lcccccc}
\hline\hline
                       &$\Delta$\teff      &$\Delta$\logg      &$\Delta$\vmicro    & $\Delta$[A/H]      & $\sigma_{\rm fit}$  &  $\sigma_{\rm total}$\\  
                       &$\pm$50~K          & $\pm$0.20         &$\pm$0.20~\kmsec   &    0.10~dex        &                    &                     \\
\hline
$\rm {[C/Fe]}$         &$\pm$0.04/$\pm$0.03&$\mp$0.01/$\pm$0.01&$\pm$0.00          &$\mp$0.02/$\pm$0.00 &$\pm$0.03/$\pm$0.01    & 0.05/0.03 \\
$\rm {[N/Fe]}$         &$\pm$0.08/$\pm$0.04&$\pm$0.01/$\mp$0.05&$\pm$0.00          &$\mp$0.02/$\pm$0.00 &$\pm$0.25/$\pm$0.08    & 0.26/0.10 \\
$\rm {[Na/Fe]}$        &$\pm$0.04          &$\mp$0.06          &$\mp$0.03/$\mp$0.05&$\pm$0.00           &$\pm$0.06/$\pm$0.03    & 0.10/0.09 \\
$\rm {[Mg/Fe]}$        &$\pm$0.03          &$\mp$0.03          &$\mp$0.02          &$\pm$0.00           &$\pm$0.04/$\pm$0.02    & 0.06/0.04 \\
$\rm {[Al/Fe]}$        &$\pm$0.00/$\pm$0.03&$\mp$0.01          &$\mp$0.01          &$\mp$0.01           &$\pm$0.05/$\pm$0.02    & 0.05/0.04 \\
$\rm {[Si/Fe]}$        &$\pm$0.01          &$\mp$0.01/$\pm$0.02&$\mp$0.03          &$\mp$0.01           &$\pm$0.06/$\pm$0.04    & 0.07/0.06 \\
$\rm {[Fe/H]}$\,{\sc i}&$\pm$0.05          &$\mp$0.01/$\pm$0.01&$\mp$0.04/$\mp$0.06&$\pm$0.00           &$\pm$0.06/$\pm$0.02    & 0.09/0.08 \\
$\rm {[Ni/Fe]}$        &$\pm$0.01          &$\pm$0.02          &$\mp$0.07          &$\pm$0.00           &$\pm$0.06              & 0.10      \\

\hline
\end{tabular}
\end{table*}

\begin{table*}
\caption{Mean chemical abundances for all the analysed SGB and RGB stars, for the subsamples of bright and faint SGBs and for the RGB stars. Average values are listed together with their associated errors, rms and the number of stars used for the means.\label{tab:average}}
\begin{tabular}{lcccc cccc cccc cccc}
\hline\hline
                        & average          & rms & \#  && average     & rms        & \#  &&average     & rms        & \#&&average     & rms        & \# \\ 
                        & \multicolumn{3}{c}{all stars}&&\multicolumn{3}{c}{bright-SGB}&&\multicolumn{3}{c}{faint-SGB}&&\multicolumn{3}{c}{RGB} \\ \hline       
$\rm {[C/Fe]}$          & $-$0.26$\pm$0.01 & 0.11 & 70 && $-$0.23$\pm$0.01 & 0.09  & 40  && $-$0.36$\pm$0.02 & 0.08 & 18 && $-$0.17$\pm$0.03 & 0.09 & 12\\ 
$\rm {[N/Fe]}$          & $+$0.88$\pm$0.06 & 0.41 & 54 && $+$0.78$\pm$0.05 & 0.27  & 29  && $+$1.31$\pm$0.08 & 0.28 & 15 && $+$0.54$\pm$0.13 & 0.38 & 10\\ 
$\rm {[Na/Fe]}$         & $+$0.16$\pm$0.02 & 0.17 & 74 && $+$0.08$\pm$0.02 & 0.13  & 41  && $+$0.34$\pm$0.03 & 0.12 & 21 && $+$0.12$\pm$0.04 & 0.13 & 12 \\ 
$\rm {[Mg/Fe]}$         & $+$0.33$\pm$0.01 & 0.09 & 74 && $+$0.31$\pm$0.02 & 0.10  & 41  && $+$0.35$\pm$0.02 & 0.07 & 21 && $+$0.33$\pm$0.02 & 0.07 & 12\\ 
$\rm {[Al/Fe]}$         & $+$0.34$\pm$0.02 & 0.15 & 40 && $+$0.28$\pm$0.03 & 0.12  & 15  && $+$0.45$\pm$0.04 & 0.15 & 13 && $+$0.28$\pm$0.03 & 0.10 & 12\\ 
$\rm {[Si/Fe]}$         & $+$0.40$\pm$0.02 & 0.17 & 62 && $+$0.35$\pm$0.03 & 0.16  & 32  && $+$0.48$\pm$0.04 & 0.18 & 19 && $+$0.38$\pm$0.02 & 0.08 & 11\\  
$\rm {[Fe/H]}$\,{\sc i} & $-$0.73$\pm$0.01 & 0.09 & 74 && $-$0.73$\pm$0.01 & 0.09  & 41  && $-$0.72$\pm$0.02 & 0.09 & 21 && $-$0.77$\pm$0.02 & 0.08 & 12\\   
$\rm {[Ni/Fe]}$         & $-$0.05$\pm$0.04 & 0.13 & 11 && --               & --    & --  && --               & --   & -- && $-$0.05$\pm$0.04 & 0.13 & 11\\  
\hline
\end{tabular}
\end{table*}

\end{document}